\documentclass[preprintnumbers,prd,onecolumn,floatfix,superscriptaddress, nofootinbib]{revtex4-1}
\usepackage{graphicx}
\usepackage{epsfig}
\usepackage{bm}
\usepackage{amssymb}
\usepackage{float}
\usepackage{amsmath}
\usepackage{dcolumn}
\usepackage[colorlinks]{hyperref}
\usepackage[usenames,dvipsnames]{color}

\hypersetup{ breaklinks=true, pdfstartview={FitH}, colorlinks=true, linkcolor=blue, citecolor=red, filecolor=magenta, urlcolor=blue, anchorcolor=green, linktocpage=true }

\def\doi{http://doi.org}

\begin{document}

\title{Cosmological aspects of a hyperbolic solution in $f(R,T)$ gravity}
\author{Ritika Nagpal}
\email{ritikanagpal.math@gmail.com}
\affiliation{Department of Mathematics, Netaji Subhas Institute of Technology, Faculty of Technology, University of Delhi, New Delhi-110 078, India}
\author{J. K. Singh}
\email{jainendrrakumar@rediffmail.com}
\affiliation{Department of Mathematics, Netaji Subhas Institute of Technology, Faculty of Technology, University of Delhi, New Delhi-110 078, India}
\author{A. Beesham}
\email{beeshama@unizulu.ac.za}
\affiliation{Department of Mathematical Sciences, University of Zululand, Kwa-Dlangezwa 3886, South Africa}
\author{Hamid Shabani}
\email{h.shabani@phys.usb.ac.ir}
\affiliation{Physics Department, Faculty of Sciences,  University of Sistan and Baluchestan, Zahedan, Iran}

\begin{abstract}
This article deals with a cosmological scenario in $ f(R,T) $ gravity for a flat FLRW model of the universe. We consider the $ f(R,T) $ function as $ f(R)+f(T) $ which starts with a quadratic correction of the geometric term $ f(R) $ having structure $ f(R)=R+\alpha R^2 $, and a linear matter term $ f(T)=2\lambda T $. To achieve the solution of the gravitational field equations in the $ f(R,T) $ formalism, we take the form of a geometrical parameter, \textit{i.e.} scale factor $ a(t)= sinh^{\frac{1}{n}}(\beta t) $ \cite{cha}, where $ \beta $ and $ n $ are model parameters. An eternal acceleration can be predicted by the model for $ 0<n<1 $, while the cosmic transition from the early decelerated phase to the present accelerated epoch can be anticipated for $ n\geq 1 $. The obtained model facilitate the formation of  structure in the Universe according to the Jeans instability condition as our model transits from radiation dominated era to matter dominated era. We study the varying role of the equation of state parameter $ \omega $. We analyze our model by studying the behavior of the scalar field and discuss the energy conditions on our achieved solution. We examine the validity of our model via Jerk parameter, Om diagnostic, Velocity of sound and Statefinder diagnostic tools. We investigate the constraints on the model parameter $ n $ and $ H_0 $ (Hubble constant) using some observational datasets: $SNeIa$ dataset,  $ H(z)$ (Hubble parameter) dataset, $ BAO $ (Baryon Acoustic Oscillation data) and their combinations as joint observational datasets $ H(z)$ + $ SNeIa $ and $ H(z)$ + $ SNeIa $ + $ BAO $. It is testified that the present study is well consistent with these observations. We also perform some cosmological tests and a detailed discussion of the model.\\
\end{abstract}

\maketitle


\section{ Introduction}

\noindent \qquad  Einstein field equations (EFE) are
\begin{equation}\label{1}
R_{ij}-\frac{1}{2} R g_{ij}+ \Lambda g_{ij} = \frac{8\pi G}{c^4}T_{ij}.
\end{equation}
In the above equation, $ R_{ij} $ is the Ricci tensor, $R$ the Ricci scalar, $ g_{ij} $  the covariant metric tensor of order 2, $ \Lambda $ the cosmological constant, $ G $ the gravitational constant, $ c $ indicates the speed of the light, and $ T_{ij} $  the energy-momentum-tensor (EMT). Despite the fact that general relativity (GR) is extremely well tested, alternatives are always present. According to observations, $ 95\% $ of the matter content of the Universe is unexplored.  GR has several problems, as the problem of initial big-bang spacetime singularity \cite{car,ein,yil} and it is not yet  quantised. GR has to be reconciled with quantum physics to discuss quantum effects. GR together with quantum physics forms the backbone of modern physics. The $ \Lambda $CDM model of the cosmology is quite successful, but there remain several unresolved issues such as the fine tuning problem \cite{bull}. Hence it is worthwhile to examine alternative theories of gravity. \\

Amongst large range of modified theories of gravity, $f(R)$ gravity \cite{tho} is considered as an interesting alternative. A more general function of $R$ \textit{i.e.} $ f(R)$ is considered in the Einstein-Hilbert action . An intensively study on $ f(R)$ gravity seems to indicate that it is an improvement over GR \cite{suj,noji,cap1}. It can also explain both phases of cosmic acceleration even in the absence of $\Lambda$ (early and late times) \cite{paul}. $ f(R) $ gravity behaves extremely well on large scales, but the theory does not hold good on all observational tests, such as on rotation of curved spiral galaxies \cite{chi1, olmo1} and the solar system regime \cite{eri,olmo2}. The generalisation of $f(R)$ gravity to $f(R, S_m)$, where the matter Lagrangian $S_m$ is considered as a general function of trace $ T $ of the EMT is termed $ f(R,T) $ gravity. Some solar system tests \cite{myr1,mor2} have been favorably applied to modified $f(R,T)$ theory of gravity to resolve the above-mentioned issues. To introduce exotic imperfect fluids and quantum effects, a trace $ T $  dependent term is determined. Generally, the source term is a function of the matter Lagrangian $ S_m $, which yields an explicit set of field equations. A lot of remarkable work in  cosmology and astrophysics has already been done in $f(R,T)$ gravity by several authors \cite{shab2, alv,sin1, shab3, shab5, sin2, sin3, sin7} and also this theory can resolve the dark matter issue \cite{sin4, zar, sin5}.\\

The above study has prompted us to compose a cosmological scenario within $ f(R,T) $ gravity. The paper is arranged as follows. In Sect. II, we give a concise discussion on $f(R,T)$ theory. We obtain highly non-linear field equations by considering the $ f(R,T) $ function as the combination of a quadratic R-dependent term and a linear T-dependent term. To determine the solution of the field equations, we use an ansatz for the scale factor $a(t)$, and find the behavior of the other geometrical parameters $ H(t) $, $ q(t) $. We also present the graphical behavior of $ \rho $, $ p $ and $ \omega $ for the obtained model in Sect. III. Next Sect. IV is devoted to the analysis and interpretation of the obtained solution by examining the potential of the scalar field and energy conditions. In Sect. V, we perform some observational tests using Jerk parameter, Om diagnostic, Velocity of sound and Statefinder diagnostic tools to explore the validity of our model. In Sect. VI, we observe that present study is well behaved with some observational datasets. In Sect. VII, some cosmological tests are discussed to calculate distances in cosmology for the accepted parametrization. Finally, we summarize our results by providing a brief conclusion about the work in Sect. VIII.

\section{ Review of f(R,T)=f(R)+2f(T) cosmology}

\qquad The general action for  $ f(R,T)=f(R)+2f(T) $  gravity \cite{har} coupled with the action of a matter field with matter Lagrangian $ S_m $ reads
\begin{equation} \label{2}
S= \int\Big(\frac{1}{16\pi G} f(R,T)+S_m\Big)\sqrt{-g} dx^4,
\end{equation}
where $ f(R,T) $ being an arbitrary function of $ R $ and $ T $. Here, we consider $ f(R)=R+\alpha R^2 $ which is the first model for inflation and it was proposed by Starobinsky \cite{star1}. This form of $ f(R) $ function takes its origin in the quantum correction to Friedmann equations. The term $ R^2 $ appears in the functional form of $ f(R) $ is the natural correction to GR and it naturally provides an inflationary scenario in early Universe. Also Starobinsky model shows the best compatibility according to the latest observations of the Universe \cite{pla} and this model serves as a possible substitute to the scalar field models describing inflation \cite{jdb}. Consequently, if one extend the above assumed $ f(R) $ form with negative exponents of curvature term, then this model is able to express the recent accelerating expansion. Therefore, with the most general model $ f(R)= R+ \alpha R^m+ \beta \frac{1}{R^n} $, where $ \alpha $ and $ \beta $ are arbitrary constants, both the acceleration in the Universe (early and late time acceleration) can be explained by the theories beyond GR \cite{paul}. To introduce exotic imperfect fluids and  taking quantum effects in to account with the above-defined $ f(R) $ model, a trace $T$ dependent term is much needed. This source term is a function of matter Lagrangian $ S_m $ which yields an explicit set of field equations. Here, in this study, we assume $ f(T) $ as a linear function of $ T $ defined as $ f(T)=2\lambda T $. So the complete form of $ f(R,T) $ function is $ R+\alpha R^2+2\lambda T $. \\

On defining EMT of matter \cite{har}
\begin{equation}\label{3}
T_{ij}= -\frac{2}{\sqrt{-g}} \frac{\delta(\sqrt{-g}S_m)}{\delta g^{ij}},
\end{equation}
where its trace is given by $T=g^{ij}T_{ij}$. Also if $S_m$ is dependent only on $g_{ij}$, in that case one can write
\begin{equation}\label{4}
T_{ij}= g_{ij}S_m-2\frac{\delta S_m}{\delta g^{ij}}.
\end{equation}
Taking a variation of action (\ref{2}) \textit{w.r.t.} $g_{ij}$, we have
\begin{equation}\label{5}
f^{R}(R,T) R_{ij}-\frac{1}{2} g_{ij} f(R,T)+(g_{ij}\Box-\nabla_i \nabla_j)f^{R}(R,T)=8\pi G\, T_{ij}-f^{T}(R,T)(T_{ij}+\Theta_{ij}),
\end{equation}
where $f^{R}(R,T)$ and $f^{T}(R,T)$ represent the derivative of $f(R,T)$ \textit{w.r.t.} $R$ and $T$ respectively, $\Box$ is the d' Alembert operator defined by $\Box= g^{ij}\nabla_{i} \nabla_{j}$ and $\nabla_i$ indicates the covariant derivative \textit{w.r.t.} $g_{ij}$ associated with the symmetric Levi-Civita connection.
$\Theta_{ij}$ is of the following form
\begin{equation}\label{6}
\Theta_{ij}\equiv g^{lm} \frac{\delta T_{lm}}{\delta g^{ij}}= -2T_{ij}+g_{ij}S_m-2 g^{lm} \frac{\delta^2 S_m}{\delta g_{ij} \delta g^{lm}}.
\end{equation}
We consider perfect fluid in the thermodynamic equilibrium, so in this way, in present study, one can simply set the matter Lagrangian $S_m = -p$ and we take EMT of matter as

\begin{equation}\label{6a}
T_{ij}=(\rho +p)u_{i}u_{j}-p g_{ij},
\end{equation}%
where $ \rho $ is the energy density and $ p $ is the pressure of the fluid present in the Universe. Using (\ref{6}), the expression for the variation of EMT of perfect fluid is given by
\begin{equation}\label{7}
\Theta_{ij}=-2T_{ij}-p g_{ij}.
\end{equation}

Using Eq. (\ref{7}) in Eq. (\ref{5}), we get the gravitational equation of motion as
\begin{equation}\label{8}
f^{R}(R,T) R_{ij}-\frac{1}{2}g_{ij} f(R,T)+(g_{ij}\Box-\nabla_i \nabla_j) f^{R}(R,T)= 8\pi G T_{ij}+f^{T}(R,T)(T_{ij}+p g_{ij}).
\end{equation}
The connection between Ricci Scalar $R$ and $T$ can be seen by contracting the Eq. (\ref{8}) \textit{w.r.t} $g^{ij}$,
\begin{equation}\label{9}
R f^{R}(R,T)- 2 f(R,T)+3\Box f^{R}(R,T) = 8\pi G\, T+(T+4p) f^{T}(R,T).
\end{equation}
On rearranging the terms in Eq. (\ref{7}), the Ricci tensor $R_{ij}$ takes the form
\begin{equation}\label{10}
R_{ij}=\frac{1}{f^{R}(R,T)}\Big(8\pi G\,T_{ij}+\frac{1}{2}g_{ij}+(\nabla_i \nabla_j -g_{ij} \Box)f^{R}(R,T)+f^{T}(R,T)(T_{ij}+p g_{ij}) \Big).
\end{equation}
Let us define a new operator $\Diamond_{ij}$ as,
\begin{equation}\label{11}
\Diamond_{ij}=\nabla_{i} \nabla_{j} -g_{ij} \Box.
\end{equation}
So the Eq. (\ref{10}) becomes
\begin{equation}\label{12}
R_{ij}=\frac{1}{f^{R}(R,T)}\Big(8\pi G\,T_{ij}+\frac{1}{2}g_{ij}+\Diamond_{ij}f^{R}(R,T)+f^{T}(R,T)(T_{ij}+p g_{ij}) \Big).
\end{equation}
The expression for  Ricci scalar $R$ can be written by arranging the terms in Eq. (\ref{9})
\begin{equation}\label{13}
R= \frac{1}{f^{R}(R,T)}\Big(8\pi G\,T+2f(R,T)-3\Box f^{R}(R,T)+(T+4p) f^{T}(R,T) \Big).
\end{equation}

By using Eqs. (\ref{12}) and (\ref{13}), the Eq. (\ref{8}) can be represented as the field equations with LHS as the Einstein tensor $G_{ij}$,
\begin{eqnarray}\label{14}
G_{ij}=R_{ij}-\frac{1}{2}R g_{ij}&=&\frac{8\pi G\, T_{ij}}{f^{R}(R,T)}+\frac{1}{f^{R}(R,T)}\Big[\frac{1}{2} g_{ij}(f(R,T)-R f^{R}(R,T))+\Diamond_{ij} f^{R}(R,T)+(T_{ij}+pg_{ij})f^{T}(R,T)\Big],\\
&=& \frac{8\pi G}{f^{R}(R,T)}(T_{ij}+T_{ij}^{'})\nonumber,
\end{eqnarray}
where $T_{ij}^{'}=\frac{1}{8\pi G}\Big(\frac{1}{2} g_{ij}(f(R,T)-R f^{R}(R,T))+\Diamond_{ij} f^{R}(R,T)+(T_{ij}+pg_{ij})f^{T}(R,T)\Big).$ From the above field equations, EFE in GR can be resumed by fixing $\alpha=0$ and $\lambda=0$. Applying the Bianchi identity on Eq. (\ref{14}) leads to\footnote{Note that this equation has been obtained in~\cite{shab4}. However, because of the metric signature in the present work, the last term in Eq. (\ref{145}) has obtained the opposite sign.}
\begin{align}\label{145}
\Big(8\pi G +f^{T}(R,T)\Big)\nabla^{i}T_{ij}+\frac{1}{2}f^{T}(R,T)\nabla_{i}T +T_{ij}\nabla^{i}f^{T}(R,T)+\nabla_{j}\Big(pf^{T}(R,T)\Big)=0.
\end{align}
\section{Cosmological dynamics of the Universe}
\qquad We study the dynamics of the Universe by considering a homogeneous and isotropic Universe in the form of spatially flat FLRW line element given by
\begin{equation}\label{15}
ds^{2}=dt^{2}-a^{2}(t)(dx^{2}+dy^{2}+dz^{2}),
\end{equation}%
where $a(t)$ being the scale factor. 
The trace $ T $ of the EMT (\ref{6a}) and scalar curvature $ R $ are
\begin{equation}\label{17}
T=\rho -3p,
\end{equation}
\begin{equation}\label{18}
R=-6(2H^2+\dot{H}),
\end{equation}
where $ H$ is the Hubble parameter defined as $\frac{\dot{a}}{a} $ and overhead dot indicates the differentiation \textit{w.r.t.} to $ t $. Taking $ f(R,T)=R+\alpha R^2 +2\lambda T $ and using Eqs. (\ref{6a}), (\ref{17}), (\ref{18}) in Eq. (\ref{14}), we get the following field equations
\begin{equation}\label{19}
3H^2=\frac{1}{1+2\alpha R}\Big[8\pi \rho+\lambda(3\rho-p)+2\alpha U(a,\dot{a},\ddot{a},\dddot{a})\Big],
\end{equation}
\begin{equation}\label{20}
2\dot{H}+3H^2=\frac{1}{1+2\alpha R}\Big[-8\pi p+\lambda(\rho-3p)+2\alpha V(a,\dot{a},\ddot{a},\dddot{a},\ddddot{a})\Big],
\end{equation}
where $U(a,\dot{a},\ddot{a},\dddot{a})=\frac{-9}{a^4}(5\dot{a}^4+a\ddot{a}^2-2a^2\dot{a}\,\dddot{a})$ and $V(a,\dot{a},\ddot{a},\dddot{a},\ddddot{a})=\frac{3}{a^4}(\dot{a}^4-18a\dot{a}^2\ddot{a}+4a^2\dot{a}\,\dddot{a}+a^2(-2\dot{a}\dddot{a}+a\ddddot{a}))$ are the functions of scale factor $a$ and its derivatives up to fourth order respectively. Also, we have set the units so that $G=1$.\\
 
Substituting the mentioned choice for function $f(R,T)$ in Eq. (\ref{145}) leads to
\begin{equation}\label{201}
\frac{8 \pi+3\lambda}{8 \pi+2\lambda}\dot{\rho}-\frac{\lambda}{8 \pi+2\lambda}\dot{p}+3H(\rho+p)=0.
\end{equation}
Eq. (\ref{19}) can be rewritten as
\begin{equation}\label{201-1}
\left(8\pi+3\lambda \right)\rho-\lambda p=\mathcal{U},
\end{equation}
where we have defined
\begin{equation}\label{201-2}
\mathcal{U}\equiv3H^{2}+18\alpha\left(\dot{H}^{2}-4H^{2}\dot{H}-2H\ddot{H}\right).
\end{equation}
Solving Eqs. (\ref{201}) and (\ref{201-1}) gives 
\begin{align}
&\rho=\frac{3 \mathcal{U}-\lambda\left(8\pi+2 \lambda \right)^{-1}H^{-1}\dot{\mathcal{U}}}{3 \left(8\pi+4 \lambda \right)}\label{201-3},\\
&p=\frac{\left(8\pi+3 \lambda \right)\rho-\mathcal{U}}{\lambda }\label{201-4}.
\end{align}

Solutions (\ref{201-3}) and (\ref{201-4}) show that to obtain the exact solutions for $ \rho $ and $ p $ and to study dark energy model, we need to adopt a parametrization of either  $ a(t) $ or $ H(t) $. This technique is called \textit{model independent way} to explore dark energy models. This work deals with an ad hoc choice of $ a(t) $, which is the outcome of a time-dependent deceleration parameter (DP) \cite{cha} as

\begin{equation}\label{35}
a(t)= \sinh^{\frac{1}{n}}(\beta t),
\end{equation}
where $ \beta $ and $ n $, $ n>0 $ are arbitrary constants.\\

The Hubble parameter $ H(t) $ and DP $ q(t) $ can be found from Eq. (\ref{35}) as
\begin{equation}\label{36}
H(t)= \frac{\beta \coth(\beta t)}{n},
\end{equation}
and
\begin{equation}\label{37}
q(t)= n\left[1-\tanh^2(\beta t)\right]-1.
\end{equation}

In the present study, we are curious to examine the different regimes of the Universe \textit{i.e.} the phase transition from decelerated to accelerated expansion by constraining a model parameter $ n $. From Eq. (\ref{37}), DP depends on $ t $ and inflation in the Universe depends on the sign of $ q $. A positive $q$ refers the decelerating expansion while a negative $ q $ corresponds to accelerating phase of the model. For the above parametrization of $ a(t) $, our model entirely accelerates and decelerates according as $ t<\frac{1}{\beta} \tanh^{-1}(1-\frac{1}{n})^{\frac{1}{2}} $ and $ t>\frac{1}{\beta} \tanh^{-1}(1-\frac{1}{n})^{\frac{1}{2}} $ respectively, and it predicts phase transitions \textit{i.e.} $ q=0 $ when $ t=\frac{1}{\beta}\,\tanh^{-1}(1-\frac{1}{n})^{\frac{1}{2}} $. As it is well acknowledged that the Universe experiences an accelerating phase in late time, so it must had a slow expansion in the past \cite{rie, per}, in such case the   parametrization of the scale factor is rational.
\begin{table}
\caption{ Dynamics of the Universe for $a(t)= sinh^{\frac{1}{n}}(\beta t)$}
\begin{center}
\label{tabparm}
\begin{tabular}{l c c c r} 
\hline\hline
\\ 
{Time ($t$)} & \,\,\,\,\, Redshift ($z$) \,\,\,  &  \,\, \,${\small a}$  \,\, \, &  \,\, \, ${\small q}$ \,\,\,   & \,\, \, \,\,\, ${\small H}$ \,\,\,\\ 
\\
\hline 
\\
${\small t\rightarrow 0}$ & ${\small z\rightarrow \infty}$ & ${\small 0}$ & $n-1$ & ${\small \infty }$ 
\\
\\
${\small t\rightarrow \infty }$ & ${\small z\rightarrow -1}$ & ${\small \infty }$ & ${\small -1}$ & \,\,\,\,\,\, finite quantity $(=\frac{\beta}{n})$
\\
\\ 
\hline\hline  
\end{tabular}    
\end{center}
\end{table}
From Eq. (\ref{37}), the model parameters $\beta$ and $n$ are related as
\begin{equation}\label{38}
\beta t_0 = \tanh^{-1}\Big(\frac{n-q_0-1}{n}\Big)^{\frac{1}{2}},
\end{equation}\\
where $t_0$ denotes the present time and $q_0$ indicates the present value of DP. Providing different values to $n$ will give rise to different values of $\beta$ on taking present value of $t$ and $q$. Here, we consider $t_0=13.8$ and $q_0=-0.54$ \cite{abd} and plot $a(t)$, $H(t)$ and $q(t)$ for various values of $ n=0.5,1,1.35,2 $.\\

Using the relation 
\begin{equation}\label{39}
 \frac{a}{a_0}=\frac{1}{1+z},
\end{equation}
where $ a_0 $ is present value of scale factor, we evaluate  $ t(z) $, $ H(z)$ and $ q(z) $  in terms of redshift $ z $ as
\begin{equation}\label{40}
t(z)=\frac{\sinh^{-1}\sqrt{\frac{n-(1+q_0)}{(z+1)^{2n}(q_0+1)}}}{\beta}.
\end{equation}
\begin{equation}\label{41}
H(z)=\frac{\beta\,\coth\left(\sinh^{-1}\sqrt{\frac{n-(q_0+1)}{(z+1)^{2n}(q_0+1)}}\right)}{n},
\end{equation}
\begin{equation}\label{42}
q(z)=n-1-n\Bigg[\tanh\Bigg(\sinh^{-1}\sqrt{\frac{n-(1+q_0)}{(z+1)^{2n}(q_0+1)}} \Bigg)\Bigg]^2.
\end{equation}\\
The graphs of scale factor $ a $, Hubble parameter $ H $ and DP $ q $ \textit{w.r.t} $ z $ are shown as:\\
\begin{figure}[tbph]
\begin{center}
$%
\begin{array}{c@{\hspace{.1in}}cc}
\includegraphics[width=2.2 in, height=1.8 in]{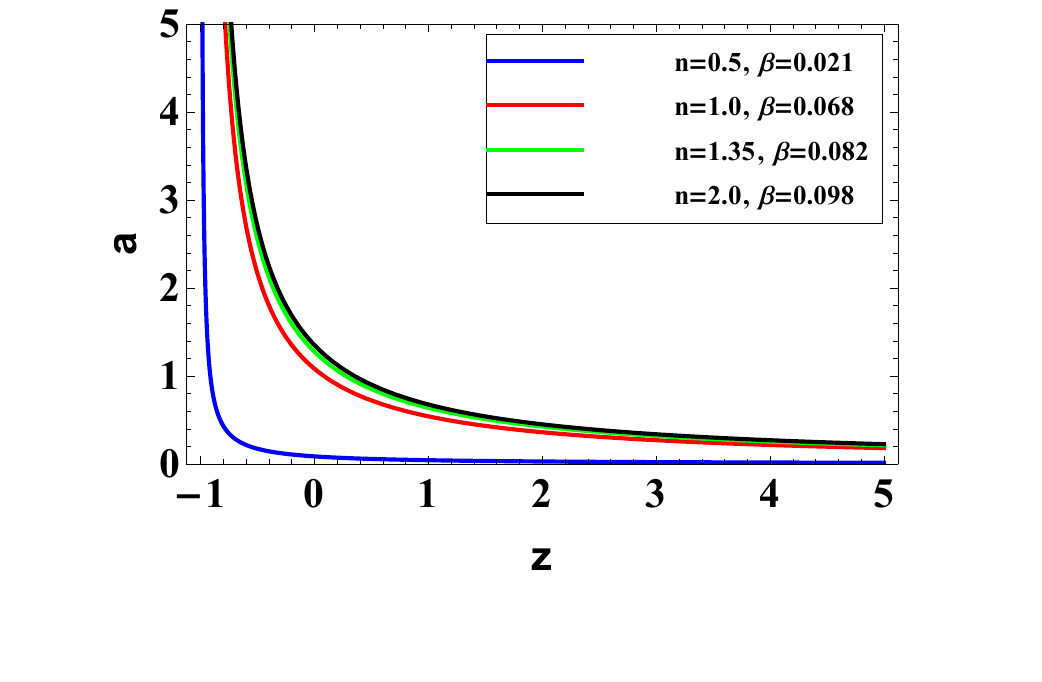} & %
\includegraphics[width=2.2 in, height=1.8 in]{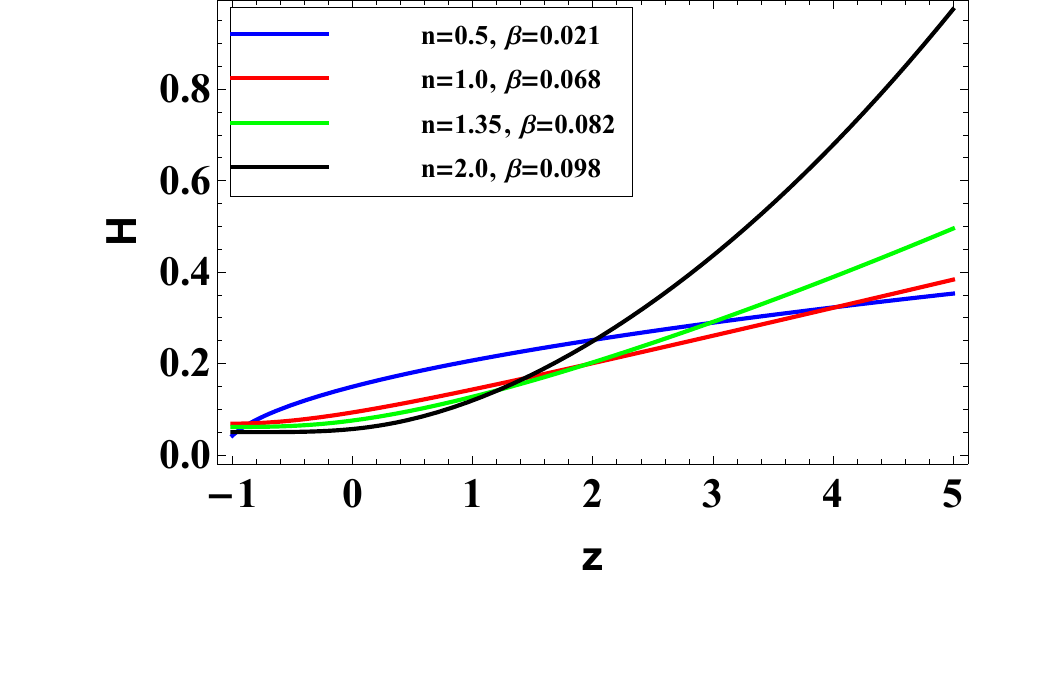} &
\includegraphics[width=2.2 in, height=1.8 in]{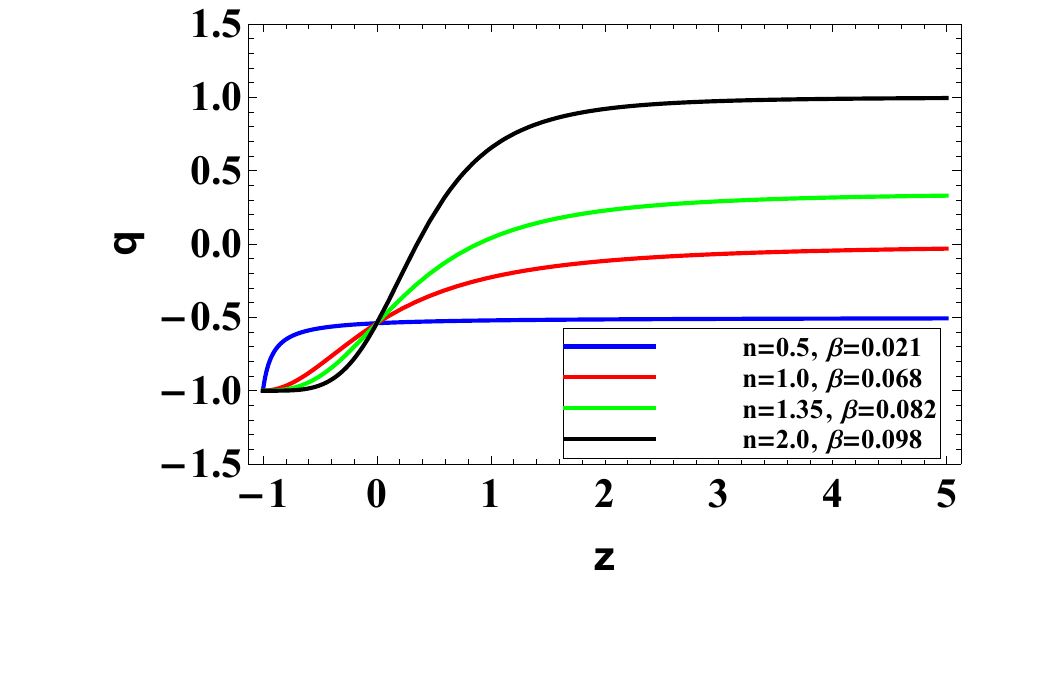}  \\
\mbox (a) & \mbox (b) & \mbox (c)
\end{array}%
$%
\end{center}
\caption{\scriptsize Graphical representations of $a$, $H$ and $q$ Vs. $ z $.}
\end{figure}

\begin{table}
\caption{ Existence of various substances according to EoS parameter}
\begin{center}
\label{tabparm}
\begin{tabular}{l c c r} 
\hline\hline
\\ 
{\textbf{Substance}} & \,\,\,\,\, \textbf{EoS parameter} \,\,\,  &  \,\, \, \textbf{Observations}
\\ 
\\
\hline 
\\
Pressureless (Cold) matter & $\omega=0$ & 32\% of the Universe
\\
\\
Hot matter & $\omega \in (0,\frac{1}{3})$ & Insignificant at present time
\\
\\
Radiation & $\omega=\frac{1}{3}$ & Influential in past
\\
\\
Hard Universe & $\omega \in (\frac{1}{3},1)$ &  Excessive high densities
\\
\\
Stiff matter & $\omega=1$ & \\
\\
\\
Ekpyrotic matter & $\omega>1$ & Resist Dominant Energy Condition
\\
\\
Quintessence & $\omega \in (0,-1)$ & 68\% of the Universe
\\
\\ 
Cosmological constant & $\omega=-1$ &  Inconsistent with observations 
\\
\\ 
Phantom Universe & $\omega<-1$ & \,\,\,\,\,Lead to Big Rip, resist Weak Energy Condition
\\
\\ 
\hline\hline  
\end{tabular}    
\end{center}
\end{table}

From Fig. 1(c), we observe that phase transitions occur when $ n\geq 1 $ and the model shows eternal acceleration when $ 0<n< 1$ \textit{i.e.} the phase transitions \textit{w.r.t} redshift $ z $ directly depend on the value of $ n $. Also Fig. 1(c) clearly specifies that our model is consistent with the recent observational dataset of $ SNeIa $, $ BAO $ and $CMBR$ (Cosmic Microwave Background Radiation) with some fine tuning, corresponding to the value of model parameter $ N=4 $ \cite{abd} when $ n=1.35 $, $ q=0 $ at $ z_{tr}=0.883752 $, and is supportive with the fitting result of Gold SNIa for $ N=182 $ with $ 1\sigma $ errors \cite{zha} when $n=2$, $q=0$ at $z_{tr}=0.352666$. Thus we can predict that the Universe started from decelerating phase and ended up with accelerating phase in late times in the case when $ n\geq 1 $ while the model represents total eternal acceleration right from the evolution of the Universe upto late time when $ 0<n<1 $. \\

The EoS parameter $\omega$ is considered as one of the vital parameter in cosmology, which explains the different cosmic regimes.  In a more generic way, this parameter can be defined as $\omega=p/\rho$, where in the case of solutions (\ref{201-3}) and (\ref{201-4}) one obtains
\begin{equation}\label{43}
\omega=-\frac{\left(8\pi+3 \lambda \right)\dot{\mathcal{U}}+3\left(8\pi+2 \lambda \right) H \mathcal{U}}{3\left(8\pi+2 \lambda \right) H \mathcal{U}-\lambda\dot{\mathcal{U}}}.
\end{equation}

In GR from the Friedmann equations, it can be observed that there is only one approach to achieve accelerated expanding Universe by considering $ 1+3\omega<0 $, which can be realised for an exotic matter, which explicitly refers negative pressure as we considered $\rho$ to be positive always. The various substances present in the Universe lead to different eras of the Universe which can be seen by providing particular values to $\omega$ (see Table II).\\

\begin{figure}[tbph]
\begin{center}
$%
\begin{array}{c@{\hspace{.1in}}cc}
\includegraphics[width=2.2 in, height=1.8 in]{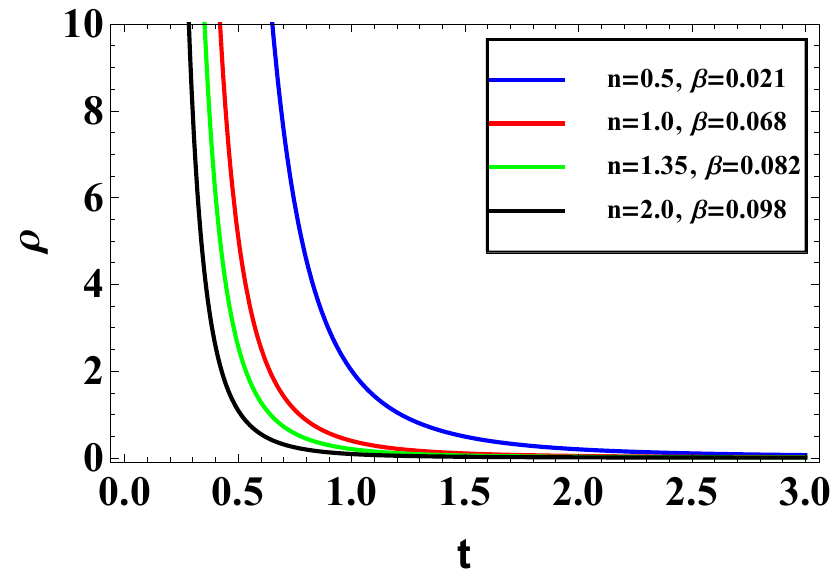} & %
\includegraphics[width=2.2 in, height=1.8 in]{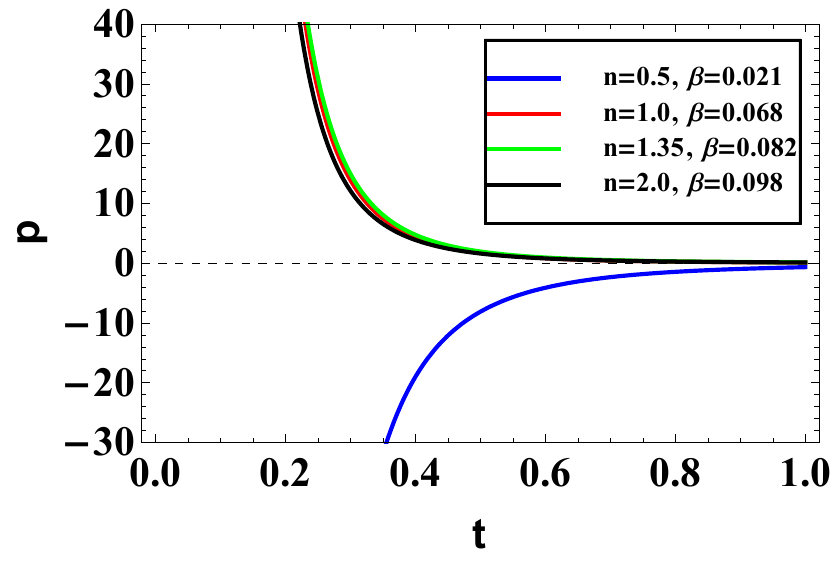}\\
\mbox (a) & \mbox (b)
\end{array}%
$%
\end{center}
\caption{\scriptsize Graphical representations of energy density $\rho$ and isotropic pressure $ p $ Vs. $ t $ for $ \alpha=0.5 $ and $ \lambda=2 $.}
\end{figure}

\qquad It is assumed that there are two major stages in the evolution of the Universe after the big bang known as the radiation and matter eras.  A radiation dominated era is requisite to anticipate primordial nucleosynthesis. Therefore, deviation of more than $ 10\% $ in expanding rate of the Universe related to the $\Lambda$CDM at the time of nucleosynthesis epoch clashes with the observed Helium abundance. Radiation and matter dominated stages are defined as the key events that help to shape the Universe. The Universe has the ability to create elements in the matter dominated era defined by the presence and pre-dominance of matter in the Universe. It features three epochs namely atomic, galactic and stellar epochs that span billion of years and includes the present day. All the three epochs are required to formulate the large structure in the Universe that we can observe today. \\

\qquad One can inspect the behavior $\rho $ and $ p $ from solutions (\ref{201-3}) and (\ref{201-4}) using the definition (\ref{201-2}). Fig. 2(a) highlights the behavior of energy density $ \rho $ which is very high \textit{i.e.} $ \rho\rightarrow \infty $ in the beginning of the Universe corresponding to $ n =0.5,\, 1,\, 1.35,\, 2 $, falls off as time unfolds and $\rho \to 0$ as $ t\to \infty $. Fig. 2(b) enacts the trait of matter pressure for all the values of $ n $ mentioned earlier. For $ n=0.5 $, $ p \to -\infty $ at $ t\to 0 $, remains negative throughout the evolution and approaches to negative constant value in late times which indicates the eternal cosmic accelerated expansion. The isotropic pressure in the early phase of the Universe for particularized values of $ n = 1,\, 1.35,\, 2 $ reaches an extensively high value and tends to $ -\frac{3 \beta ^2}{n^2 \left(8\pi+4 \lambda \right)} $ as $ z\rightarrow -1 $. Negative pressure in the universe is subjected to the acceleration in the cosmos according to the standard cosmology. Therefore, the present study exhibits accelerating phase at current epoch as well as in the near future. From Fig. 2(b), we can realize that structure formation is achievable in our model for the case $ n = 1,\, 1.35,\, 2 $ because decelerated expansion is required for the structure formation that could appear in the presence of a kind of matter fluid which produces Jeans instability \cite{james}.\\

\indent It can be useful to study the matter density and pressure in the limit of small and large times. Straightforward calculations show that both quantities tend to infinity whose signature depend upon the model constants $ \lambda $, $ \alpha $ and $ n $. In the limit of large times one obtains
\begin{equation}\label{434}
\lim_{t\to\infty}\rho=-\lim_{t\to\infty}p=\frac{3 \beta ^2}{n^2 \left(8\pi+4 \lambda \right)}.
\end{equation}
The limit value (\ref{434}) includes some interesting information. It is noted that it does not depend on the coupling constant $ \alpha $ which incorporates the curvature correction term in the Lagrangian. On the contrary, result (\ref{434}) depends on the coupling constant of the matter part,\textit{ i.e.} $ \lambda $. From (\ref{434}) we see that matter behaves like the DE in the late times. Also, value (\ref{434}) implies that to have a gravitational model with observationally accepted values for the matter density and pressure in the late times, the constraint $ \lambda>-2\pi $ must hold. More precisely, by taking the limit value of the matter density in the early times one finds that to guarantee the weak energy condition (WEC), $ \rho\geq0 $, constraints which are indicated in the Table III must hold.

\begin{table}
\caption{ Conditions to guarantee the WEC}
\begin{center}
\label{tabparm}
\begin{tabular}{l c c r} 
\hline\hline
\\ 
{} & \,\,\,\,\, $\alpha<0$ \,\,\,  &  \,\, \, $\alpha>0 $ 
\\ 
\\
\hline 
\\
$ -2 \pi <\lambda <0 $ \,\,\,\,\, & \,\,\,\,\, $ n>  -(\frac{6 \pi }{\lambda }+\frac{3}{2}) $ \,\,\,\,\, & \,\,\,\,\,$ 0<n< -(\frac{6 \pi}{\lambda }+\frac{3}{2}) $ 
\\
\\
$ \lambda>0 $ & \,\,\,\,\, & \,\,\,\,\, $ n>0 $
\\
\\ 
\hline\hline  
\end{tabular}    
\end{center}
\end{table}

It is worth discussing the behavior of the EoS parameter (\ref{43}). In the early times the EoS parameter (\ref{43}) goes to a constant value and in the late times it mimics the DE, i.e., 
\begin{align}\label{434-1}
\omega_{i}\equiv\lim_{t\to0}\omega=1+\frac{2 (\lambda +4 \pi ) (2 n-3)}{\lambda  (2 n+3)+12 \pi },
\end{align}
where the subscript `$i$' represents the values of $\omega$ in the early times. For the late times one obtains 
\begin{align}\label{434-2}
\lim_{t\to\infty}\omega=-1,
\end{align}
which is independent of the model constants. Some simple calculations show that constraining the EoS parameter as $0<\omega_{i}<1$ (which is plausible in cosmology) forces one to choose $\big((3 \lambda +12 \pi )/(6 \lambda +16 \pi)˘\big)<n<3/2$ for $\lambda>-2\pi$. Also, for $-2 \pi <\lambda <0, 3/2<n<-\big((12 \pi+3 \lambda)/2\lambda \big)$ and $\lambda> 0,~ n>3/2$, the EoS parameter can behave like ekpyrotic matter $(\omega>1)$ which resists DEC followed by matter dominated era where energy density is extensively high to radiation dominated era in the early Universe. Results (\ref{434-1}) and (\ref{434-2}) contains an important informations about different epochs of cosmic evolution. From (\ref{434-1}) we learn that the underlying gravitational model can describe two different cosmological transitions:
\begin{itemize}
  \item [(i)] for $ n=3/2 $ one obtains $ \omega^{s}_{i}=1 $. That is independent of other model parameters and only for $ n=3/2 $ the model describe a transition from a state in which stiff matter dominates in the early eras to a state which behaves like the DE in the late times.
  \item [(ii)] it is also possible a transition from a pressure-less matter dominated era with $ \omega^{p}_{i}=0 $ in the early Universe to a DE like Universe in the late times. In this case one obtains the following relation
\begin{align}\label{434-3}
n=\frac{3 (\lambda +4 \pi )}{2 (3 \lambda +8 \pi )}.
\end{align}
\end{itemize}
It has been seen that the parameters $ \beta $ and $ \alpha $ do not play any role in determining the initial and final states of cosmological evolution. It means that only modifications in the matter part of the Lagrangian are responsible for different states of the Universe. The evolution of EoS parameter $ \omega $ \textit{w.r.t.} time $ t $ as well as redshift $ z $ are represented graphically in Fig. 3(a) and Fig. 3(b) when $ n=3/2, \,\frac{3 (\lambda +4 \pi )}{2 (3 \lambda +8 \pi )} $.\\

\begin{figure}[tbph]
\begin{center}
$%
\begin{array}{c@{\hspace{.15in}}cc}
\includegraphics[width=2.2 in, height=1.8 in]{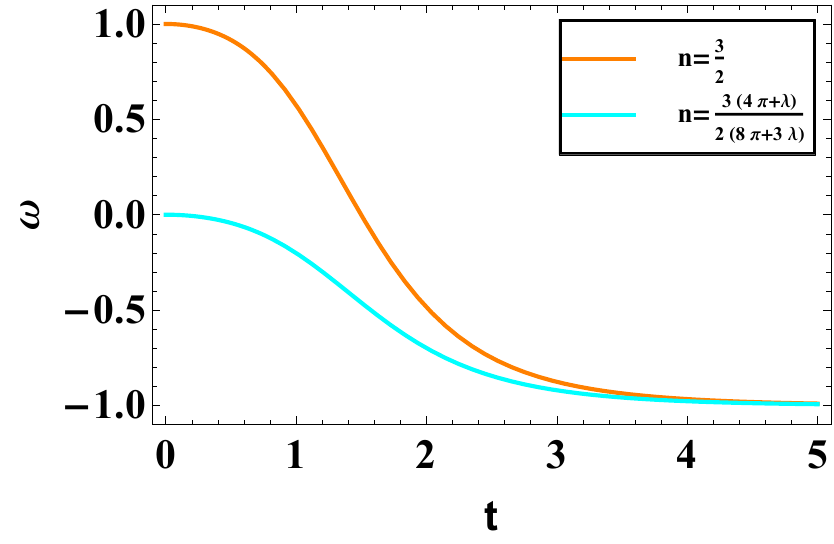}& %
\includegraphics[width=2.2 in, height=1.8 in]{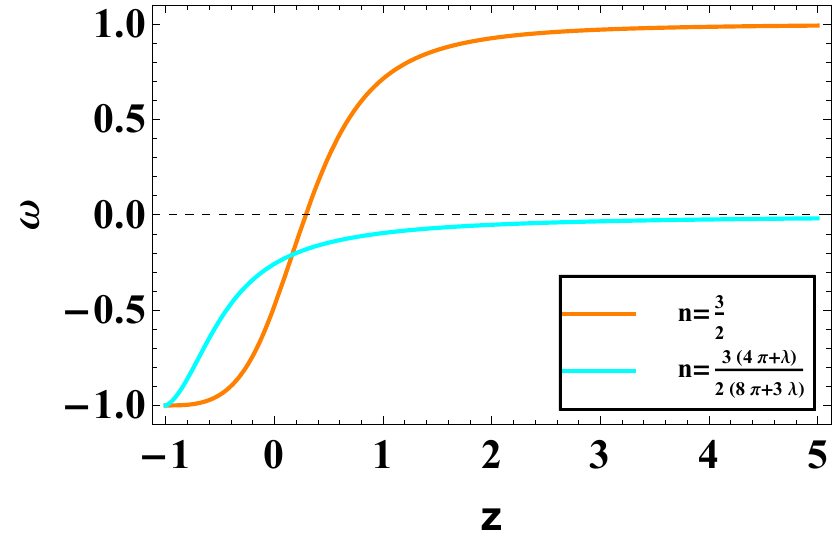}\\
\mbox(a) & \mbox(b)
\end{array}
$%
\end{center}
\caption{\scriptsize The plots of (a) EoS parameter $ \omega $  Vs.  $ t $ and (b) EoS parameter $ \omega $  Vs. $z$ when $ n=3/2, \,\frac{3 (\lambda +4 \pi )}{2 (3 \lambda +8 \pi )} $ for a fix value of $ \lambda=2 $, $ \alpha=0.5 $ and $ \beta=0.6 $.}
\end{figure}

The profound discussion on the behavior of EoS parameter which depends on the range of matter-geometry coupling constant $\lambda$ in $f(R,T)$ gravity is worthy and it has been investigated in Fig. 3. For a fix value of $ \alpha=0.5 $, and $ \lambda=2 $, $ \omega \in $ quintessence region for high redshift $ z $ and as time unfolds, $ \omega \to -1 $ in infinite future (\textit{i.e.} $ z\to -1 $) when $ n=\frac{3 (\lambda +4 \pi )}{2 (3 \lambda +8 \pi )} $, which is consistent with the observations of temperature fluctuation in cosmic microwave background radiations (CMBR) \cite{hin}. Also, the EoS parameter indicates another possibility for evolution of the Universe when $ n=3/2 $. In this case our model evolves from a state of a stiff-matter fluid dominated era to a DE like era in the late times.\\

The present value of the EoS parameter for the mentioned types of evolution era can be obtained respectively, as follows
\begin{align}\label{434-4}
\omega_{0}^{(n=3/2)}=\frac{\lambda  \left(176 \alpha  \beta ^2-9\right)+8 \pi  \left(46 \alpha  \beta ^2-9\right)}{5 \lambda  \left(32 \alpha  \beta ^2+9\right)+16 \pi  \left(19 \alpha  \beta ^2+9\right)},
\end{align}
for models with $n=3/2$ and 
\begin{align}\label{434-5}
\omega_{0}^{(n(\lambda))}=-\frac{(3 \lambda +8 \pi ) \Big[3 \lambda  \left(8 \alpha  \beta ^2+9\right)+4 \pi  \left(8 \alpha  \beta ^2+27\right)\Big]}{9 \lambda ^2 \left(88 \alpha  \beta ^2+13\right)+12 \pi  \lambda  \left(320 \alpha  \beta ^2+63\right)+128 \pi ^2 \left(35 \alpha  \beta ^2+9\right)},
\end{align}
for $ n=\frac{3 (\lambda +4 \pi )}{2 (3 \lambda +8 \pi )} $, where $\omega_{0}$ stands for the current value of the EoS parameter. From Eqs. (\ref{434-4}) and (\ref{434-5}), it can be seen, both coupling constants $\alpha$ and $\beta$ in the  matter Lagrangian affect the present value of the EoS parameter. Therefore, some suitable astronomical data can be used to constrain the values of the coupling constant. The EoS parameters $ \omega\simeq-0.509 $ and  $ \omega\simeq-0.2629 $ at present epoch $ z=0 $ for $ n=3/2, \,\frac{3 (\lambda +4 \pi )}{2 (3 \lambda +8 \pi )} $ respectively, which is in good agreement with the observation \cite{abd}.

\section{ Interpretation of the model}
\subsection{ Scalar field correspondence}
\qquad In section 3, we have already discussed that the idea to predict the acceleration in the Universe is to filled with an exotic form of matter which satisfy $1+3\omega<0$. According as the observations, the energy which produces the acceleration satisfies $\omega\simeq -1$. If $\omega<0$, there are many models that can  explain inflation exactly such as quintessence model, phantom model \textit{etc.} \cite{rat, sam, cald, sah1}. In section 3, we have also study above the construction of EoS parameter $\omega$ for our  model, so it is appropriate to consider a matter field which shows exotic behavior and is able to produce anti-gravitational effects. Here, we consider dark energy as quintessence to explain the cosmic acceleration whose action is given by 

\begin{equation}\label{44}
S= \int\Big(\frac{1}{16\pi G} R+S_m\Big)\sqrt{-g}\, dx^4,
\end{equation}
with the matter Lagrangian density 
\begin{equation}\label{45}
S_m=\int\Big(-\frac{1}{2}\partial_\mu{\phi}\,\partial^\mu{\phi}-V(\phi)\Big)\sqrt{-g}\, dx^4,
\end{equation}
where $ \phi $ is the time-dependent scalar field. Therefore, we can consider scalar field $ \phi $ as a perfect fluid with energy density $ \rho $ and pressure $ p $ as
\begin{equation}\label{46}
\rho=\frac{1}{2}\dot{\phi}^2+V(\phi),
\end{equation}

\begin{equation}\label{47}
p=\frac{1}{2}\dot{\phi}^2-V(\phi).
\end{equation}

Here $ \frac{1}{2}\dot{\phi}^2 $ is the kinetic energy $ (KE) $ and $ V(\phi) $ is the potential energy $ (PE) $ of the scalar field. So it can be noticed that $ \omega = \omega(t) $ \textit{i.e.} it can no more be treated as a constant. The quintessence or phantom model is consistent with the observations provided $ \omega\simeq -1 $. Thus, we need $ \dot{\phi}^{2}<<V(\phi) $ \textit{i.e.} the $KE$ of $\phi$ is insignificant in comparison to the $PE$. In this study, we consider that $\phi$ is the only source of DE with $V(\phi )$, so one can consider energy density and pressure of scalar field as $\rho _{\phi}$ and $p_{{\phi}}$ respectively for flat FLRW space-time under Barrow's scheme \cite{bar} using Eqs. (\ref{46}) and (\ref{47}) as
\begin{figure}[tbph]
\begin{center}
$%
\begin{array}{c@{\hspace{.1in}}cc}
\includegraphics[width=2.5 in, height=2.5 in]{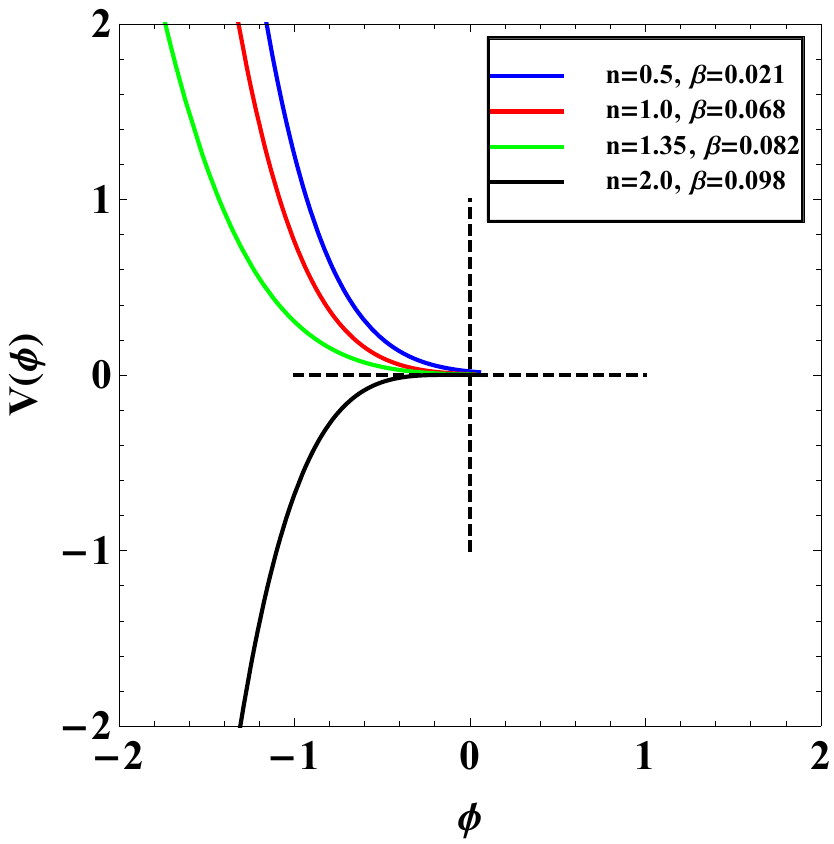}
\end{array}%
$%
\end{center}
\caption{\scriptsize The plot of potential energy $V(\phi)$ Vs. scalar field $\phi$.}
\end{figure}
\begin{equation}\label{48}
\rho=\frac{1}{2}\dot{\phi}^2+V(\phi)=\rho_{\phi},
\end{equation}
\begin{equation}\label{49}
p=\frac{1}{2}\dot{\phi}^2-V(\phi)=p_{\phi}.
\end{equation}

The $KE$ and $PE$ can be obtained by solving the Eqs. (\ref{48}) and (\ref{49}). Fig. 4 demonstrates the potential energy $ V(\phi) $ plots \textit{w.r.t.} scalar field $\phi$ for the same considered values of model parameters as we have taken in Fig. 1, 2. From Fig. 4, we notice that the potential $ V(\phi) $ is present in the interval $ -1<\phi<0 $ and $ V(\phi)\simeq0 $ at $ \phi\simeq0 $. Therefore, we can predict that the scalar field $ \phi $ is the only source of DE with potential $ V(\phi) $. Thus we conclude that our model is an accelerating dark energy model. \\

\subsection{ Energy conditions}
\qquad Energy conditions (EC) have a great utility in classical GR which discuss the singularity problems of space-time  and explain the behavior of null, space-like, time-like or light-like geodesics. It provides some extra freedom to analyse certain ideas about the nature of cosmological geometries and some relations that the stress energy momentum must satisfy to make energy positive. In general, the EC can be classified as (i) NEC (Null energy condition), (ii) WEC (Weak energy condition), (iii) SEC (Strong energy condition), and (iv) DEC (Dominant energy condition). The EC can be formulated in many ways such as geometric way (EC are well expressed in terms of Ricci tensor or Weyl tensor), physical way (EC are expressed purely by the help of stress energy momentum tensor), or effective way (EC are expressed in terms of energy density $\rho$, which serves as the time-like component and pressures $p_i,\, i=1,2,3$, which represent the $3$-space-like component). The formulation of these four types of EC in GR are point-wise expressed effectively as\\ 
\begin{itemize}
\item NEC $\Leftrightarrow$ $\rho+p_i \geq 0$, $ \forall i $,
\item WEC $\Leftrightarrow$ $\rho \geq 0$, $ \rho+p_i \geq 0 $, $ \forall i $,
\item SEC $\Leftrightarrow$ $\rho+\sum_{i=1}^3 p_i \geq 0$, $\rho+p_i \geq 0$, $ \forall i $,
\item DEC $\Leftrightarrow$ $ \rho \geq 0$ , $|p_i| \leq \rho $, $ \forall i $.
\end{itemize}
The graphical representation of NEC, SEC and DEC for a fix value of $\alpha=0.1$ and varying range of $\lambda$ are shown in Fig. 5\\
\begin{figure}[tbph]
\begin{center}
$%
\begin{array}{c@{\hspace{.1in}}cc}
\includegraphics[width=2.2 in, height=1.8 in]{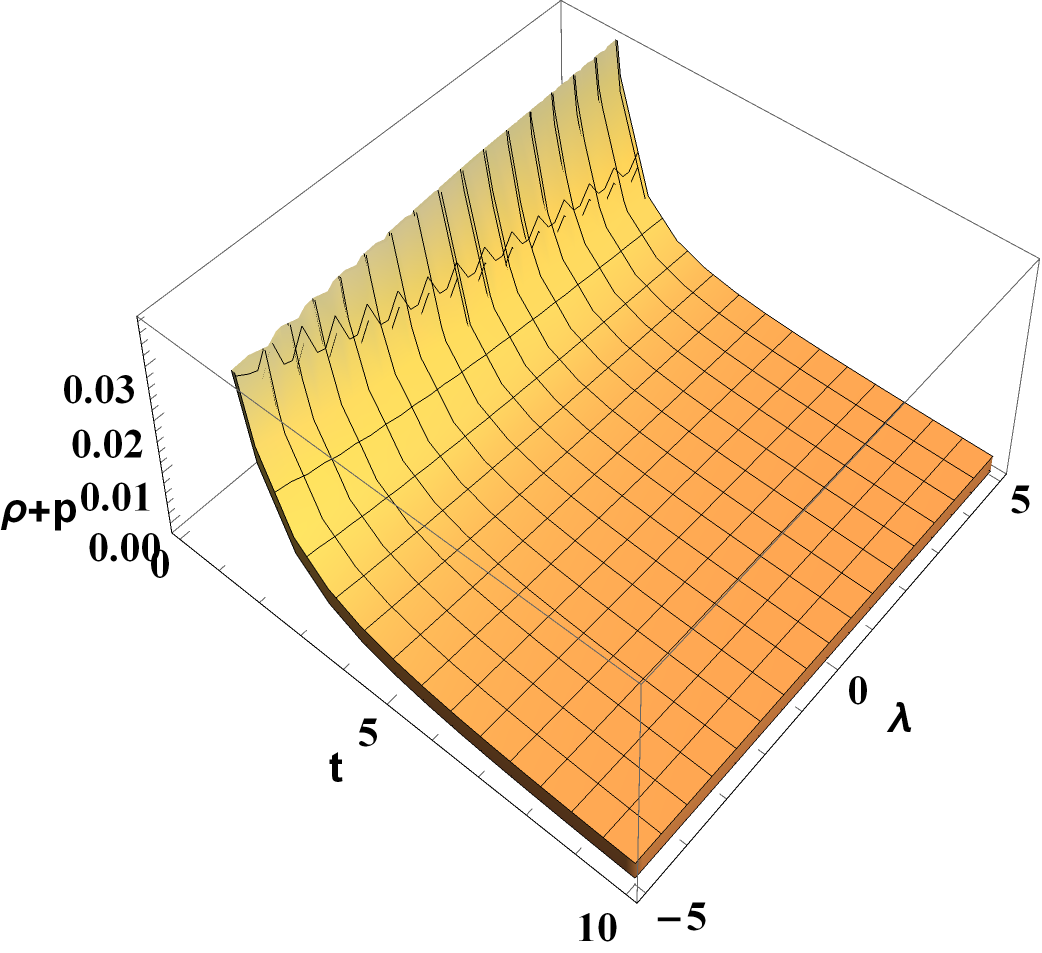} & %
\includegraphics[width=2.2 in, height=1.8 in]{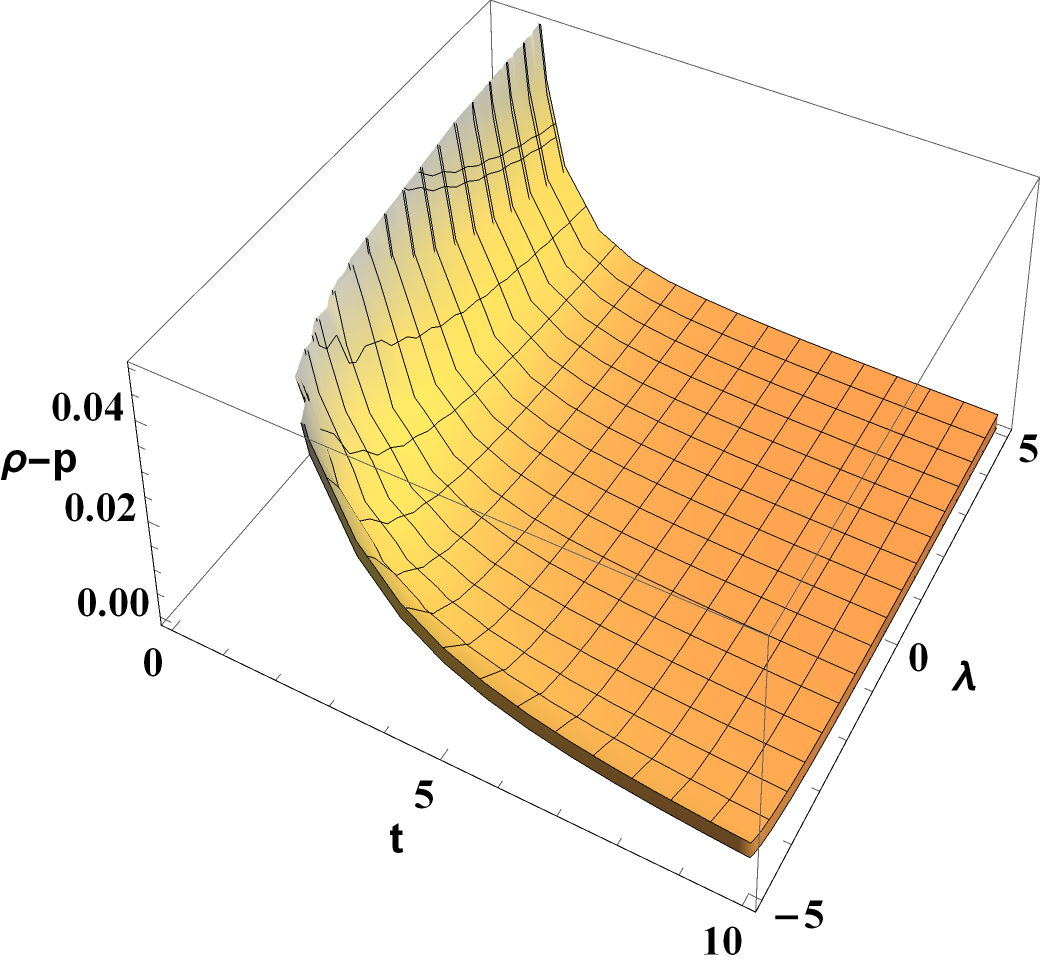}  &
\includegraphics[width=2.2 in, height=1.8 in]{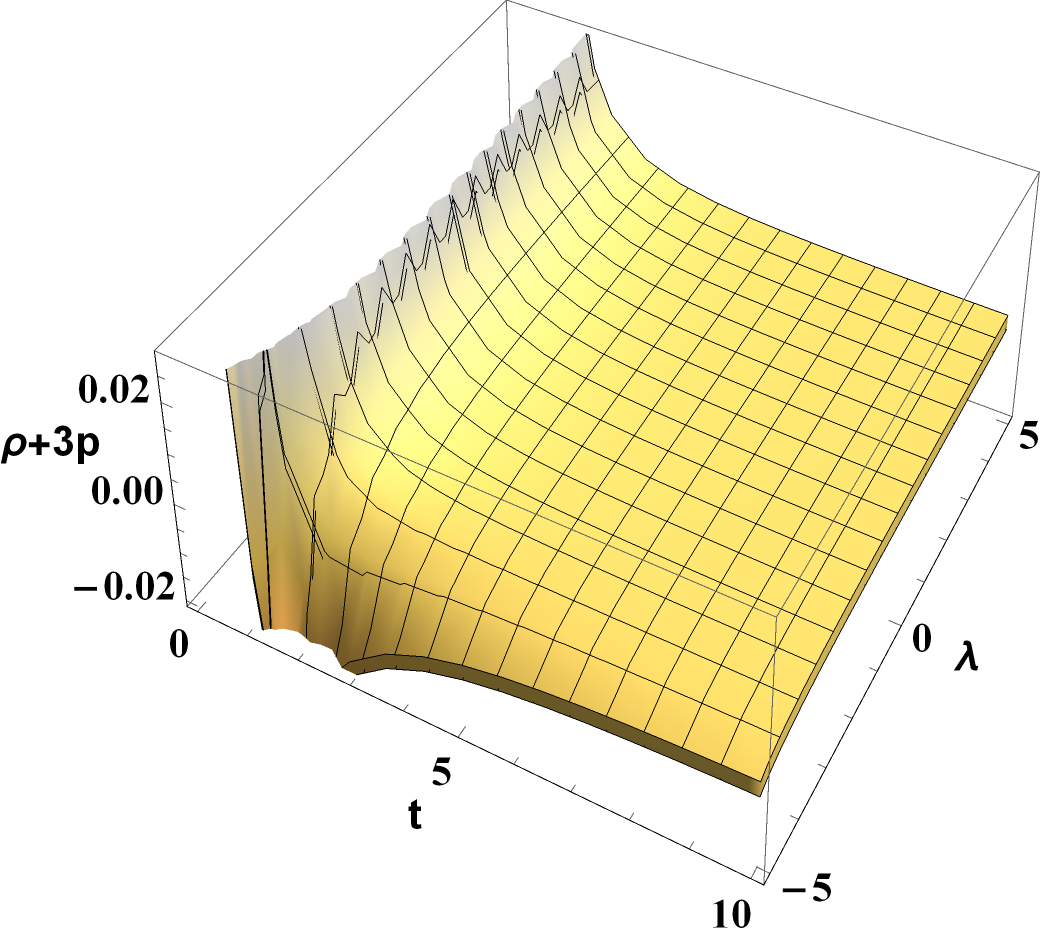} \\

\mbox (a) & \mbox (b) & \mbox (c)
\end{array}%
$%
\end{center}
\caption{\scriptsize Graphical behavior of NEC, DEC and SEC for $ n=1.35 $.}
\end{figure}

Using the above mentioned relations, we discuss all four energy conditions in $ f(R,T) $ theory for all different values of $ n=0.5,\, 1,\, 1.35,\, 2 $ and $ \alpha=0.1 $ by providing different range of coupling constant $ \lambda $. We observe the evolution of energy density and validation for all the EC for both positive and negative range of $ \lambda $. We examine that for positive $ \lambda $, NEC, WEC and DEC hold but SEC violates for $ n=0.5 $, which directly implies the accelerated expansion of the Universe. Also, as it is clear that $ \lambda $ is any arbitrary coupling constant so it can also accept the negative values, so if we extend our domain of $ \lambda $ upto negative values, then it is worth emphasizing that SEC does not hold good for all the models $ n=0.5,\, 1,\, 1.35,\, 2 $, which exactly leads to accelerating phase of the Universe. Here as a matter of discussion, we graphically sketch the figures of the energy conditions for $ n=1.35 $ only.

\section{ Validation of the model}

\subsection{ Jerk parameter}
\qquad As we know that the Hubble parameter $ H $ measures the fractional rate of change of scale factor $ a $ \textit{ i.e.} the instantaneous expansion and the second derivative of scale factor $ q $ measures the cosmic acceleration. Similarly higher derivatives of scale factors are allow us to study the cosmic expansion history and they can potentially differentiate the various dark energy models. Jerk parameter $ j $ is an extensive kinematical quantity which measures the rate of change of third derivative of scale factor \textit{w.r.t.} time $ t $. On expanding the Taylor series for scale factor around $ a_0 $, the fourth term of the Taylor series contains the jerk parameter $ j $ \cite{sin6}. The Taylor's expansion around $a_0$ containing the jerk parameter is given by 

\begin{equation}\label{50}
\frac{a}{a_0}=1+H_0t-\frac{1}{2!}q_0H_0^2t^2+\frac{1}{3!}j_0H_0^3t^3-\cdot\cdot\cdot\cdot\cdot\cdot\cdot,
\end{equation}\\
where $ H_0 $, $ q_0 $ and $ j_0 $ being the the current values of the parameters. Therefore the jerk parameter is defined as 
\begin{equation}\label{51}
j=\frac{\dddot{a}}{a H^3},
\end{equation}\\
and in terms of $q$, jerk parameter $j$ reads \cite{vis,rap}
\begin{equation}\label{52}
j=q+2 q^{2}-\frac{\dot{q}}{H}.
\end{equation}\\
Using Eqs. (\ref{35}) and (\ref{36}), it can be calculated as 

\begin{equation}\label{53}
j=1+n(2n-3)sech(\beta t)^2.
\end{equation}\\
Also it will always be suitable to express jerk parameter $ j $ in terms of redshift $ z $ when $ q(z) $ is given \cite{sah3, ala}. The Jerk parameter $ j $ in terms of $ z $ is expressed as

\begin{equation}\label{54}
j=1+\frac{n(2n-3)}{1+2.17391(n-0.46)(1+z)^{-2n}}.
\end{equation}\\
In Fig. 6(a), the cosmic jerk parameter highlights the dynamics of the Universe. Universe transits from decelerated to accelerated phase in a cosmic jerk $ j $ with a positive value $ j_0 \approx 1 $ and negative value $q_0$ in accordance with $ \Lambda $CDM. It shows the evolution of $j$ parameter for different values of $ n $ and is freely seen that $ j $ remains positive in all the cases and approaches to $1$ in late times. Jerk parameter $j$ at present ($z=0$) is positive different from $1$ in all the four cases. Therefore, we can expect another dark energy model instead of $ \Lambda $CDM.\\

\begin{figure}[tbph]
\begin{center}
$%
\begin{array}{c@{\hspace{.1in}}cc}
\includegraphics[width=2.2 in, height=2.2 in]{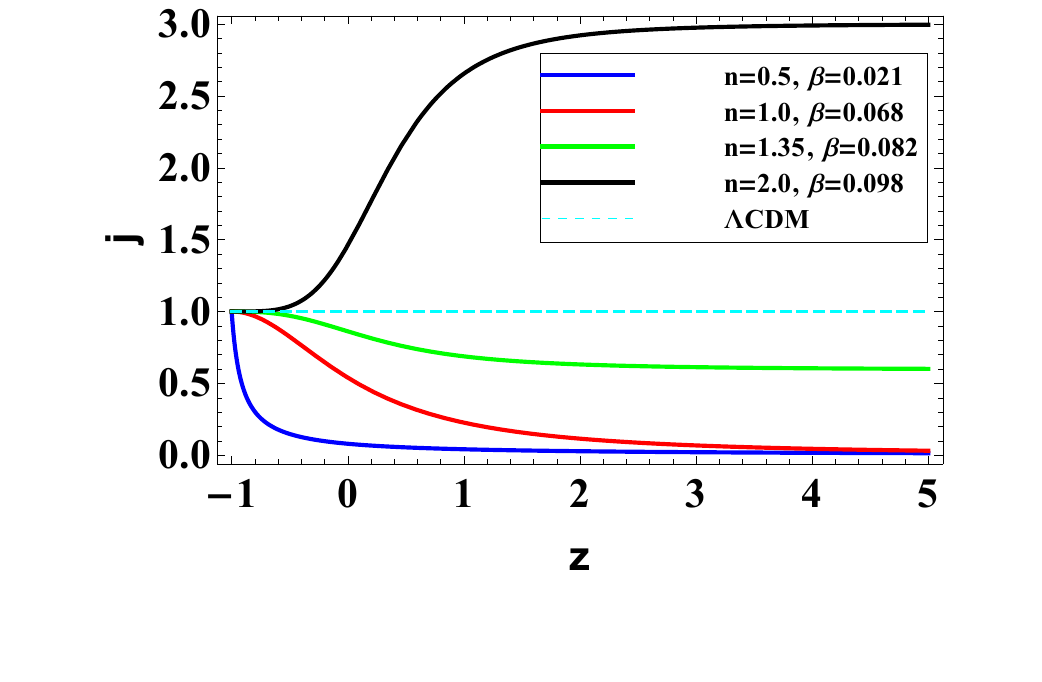}& %
\includegraphics[width=2.2 in, height=2.2 in]{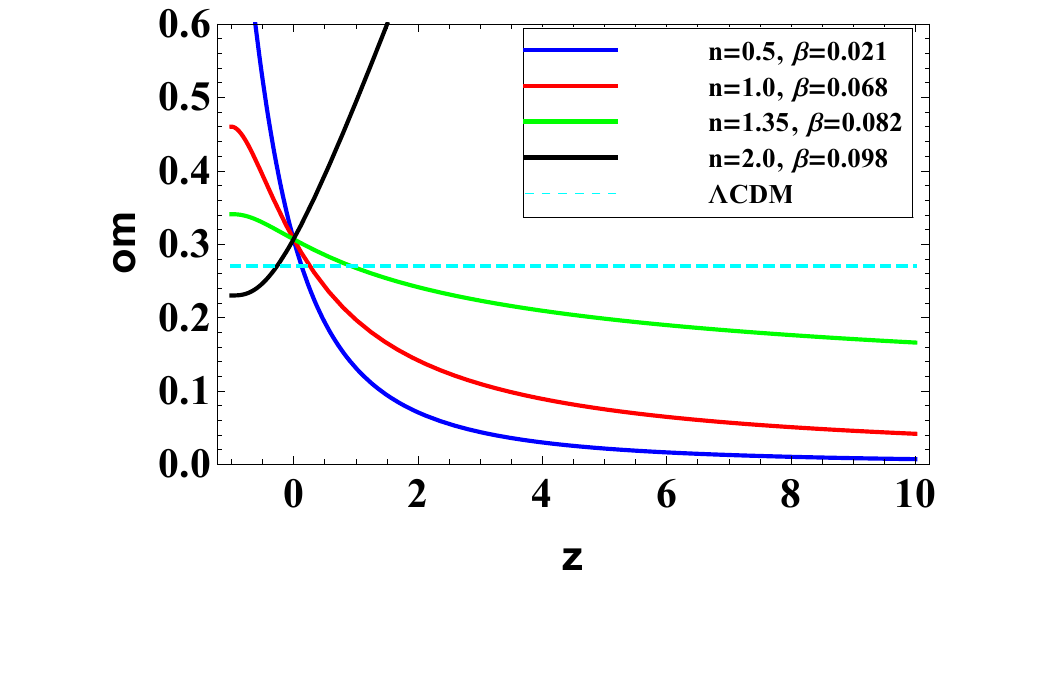}& %
\includegraphics[width=2.2 in, height=2.2 in]{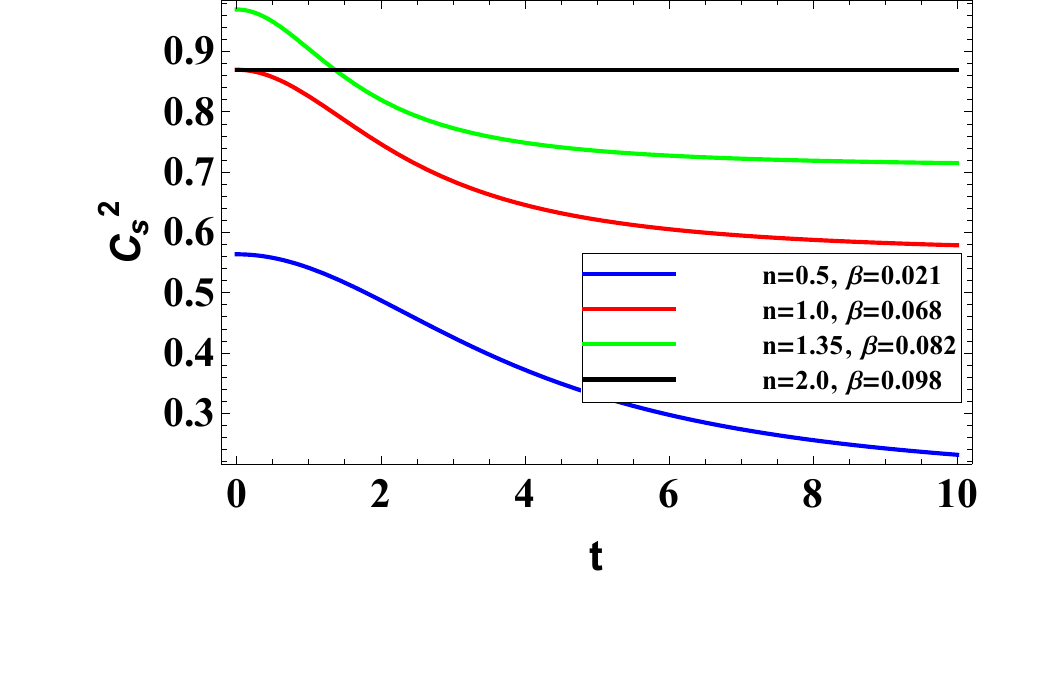}\\
 \mbox (a) & \mbox(b) & \mbox(c)
\end{array}%
$%
\end{center}
\caption{\scriptsize Graphical behavior of (a) Jerk parameter $j$ Vs. time $t$, (b) Evolution of $Om$ Vs. redshift $z$, (c) Velocity of sound $C_s^2$ Vs. time $t$.}
\end{figure}

\subsection{ Om diagnostic}
\qquad In this section we discuss the most popular diagnostic known as Om diagnostic denoted by $Om(z)$, used to distinguish standard $ \Lambda CDM $ model from various dark energy models \cite{sah4,zun}. This diagnostic is related to Hubble parameter $H$ and redshift $ z $. It is noted that different trajectories of $Om(z)$ facilitate significant differences among various DE models without actually mentioning the current value of $\Omega_{m}$ (density parameter of matter). Om diagnostic $Om(z)$ is defined as
\begin{equation}\label{55}
Om(z)=\frac{\Big(\frac{H(z)}{H_0}\Big)^2-1}{z(z^2+3z+3)}.
\end{equation}
This tool suggest a quintessence type behavior of dark energy $ (\omega>-1) $ corresponding to its negative curvature (\textit{i.e.} below the $ \Lambda CDM $ line), phantom type behavior $ (\omega<-1) $ corresponding to its positive curvature (\textit{i.e.} above the $ \Lambda CDM $ line) and $ Om(z)= \Lambda CDM $ corresponding to zero curvature. Fig. 6(b) explains the behavior of different dark energy models corresponding to different values of $n$. For $ n=0.5,\,1,\,1.35 $ model shows quintessence type behaviour $(\omega>-1)$ as graph of $Om(z)$ shows a downward trend as redshift $z$ increases and for $n=2$, model represents phantom behavior $(\omega<-1)$ as $Om(z)$ has positive slope.\\

\subsection{ Velocity of sound}
\qquad The velocity of sound is one of the stringent attempt to investigate the validity of a cosmic model. A model is said to be physically acceptable if velocity of sound $ C_{s}^{2} $ is less than the speed of light $ c $. The stability condition for the model is given by the relation $ 0\leq C_s^2=\frac{dp}{d\rho}\leq 1 $. In this study we have taken the speed of light $ c $ is 1. Therefore, the model is physically realistic provided the condition $ 0\leq\frac{dp}{d\rho}\leq 1 $ is satisfied.\\

Fig. 6(c) shows the profile of $ C_s^2 $ for $ n=0.5,\, 1,\, 1.35,\, 2 $. The constant $\lambda$ decides the stability of the model for a fix value of $ \alpha=0.1 $. The stability of the cosmic model depends on the coupling constant $ \lambda $. By considering different values of $ \lambda$, we have different stability scenarios of the model. The model satisfies the condition $ C_s^2 \leq 1 $ throughout the evolution with time $ t $ for all the cases $ n=0.5,\, 1,\, 1.35,\, 2 $ when $ \lambda $ is taken in the range $-17.5\leq \lambda \leq -12.57$. The condition $0<C_s^2<1$ does not hold for other cosmic ranges. Therefore, we can say that our model is partially stable. 

\subsection{ Statefinder diagnostic}
\qquad The present cosmic acceleration which is rational with the recent cosmological observations and the signature flipping behaviour of deceleration parameter $q$ from $+ve$ to $-ve$ in accordance with high redshift $ z $ to low redshift $ z $ enforce us to study beyond $ q $ and $ H $, and find some more cosmological models of DE other than $\Lambda$CDM. The behavior of higher derivatives of scale factor $a$ other than $H$ and $q$ are the essential components to explain the dynamics of the Universe. Due to this  reasons, we generalize our domain to construct geometrical parameters which involves higher derivatives of $a$. A technique named as Statefinder diagnostic in which a pair of geometrical parameters $\{r,s\}$  proposed by \cite{sah3, ala} is taken in to account to describe the dynamics of various DE models. These  parameters $\{r,s\}$ are defined as

\begin{equation}\label{56}
r=\frac{\dddot{a}}{aH^{3}}\text{, \ \ }s=\frac{r-1}{3(q-\frac{1}{2})},
\end{equation}%
where $q\neq \frac{1}{2}$.\\

For our parametrization of $a$ in Eq. (\ref{35}), the expressions of $ r $ and $ s $ are given as follows:

\begin{equation}\label{57}
r=1+n(2n-3)sech(\beta t)^2,
\end{equation}
where the parameter $ r $ is same as the jerk parameter $ j $, which is defined in subsection $ 5.1 $.

\begin{equation}\label{58}
s=\frac{4n(2n-3)}{-9+12n-9 cosh(2t\beta)}.
\end{equation}

This technique facilitate us that how one can differentiate various DE models easily by plotting the different trajectories of $r$ and $s$ (see Fig. 7). For a brief and recent review on statefinder diagnostic, see \cite{sin2, rit2, srt}.\\       

\begin{figure}[tbph]
\begin{center}
$%
\begin{array}{c@{\hspace{.1in}}cc}
\includegraphics[width=2.5 in, height=2.5 in]{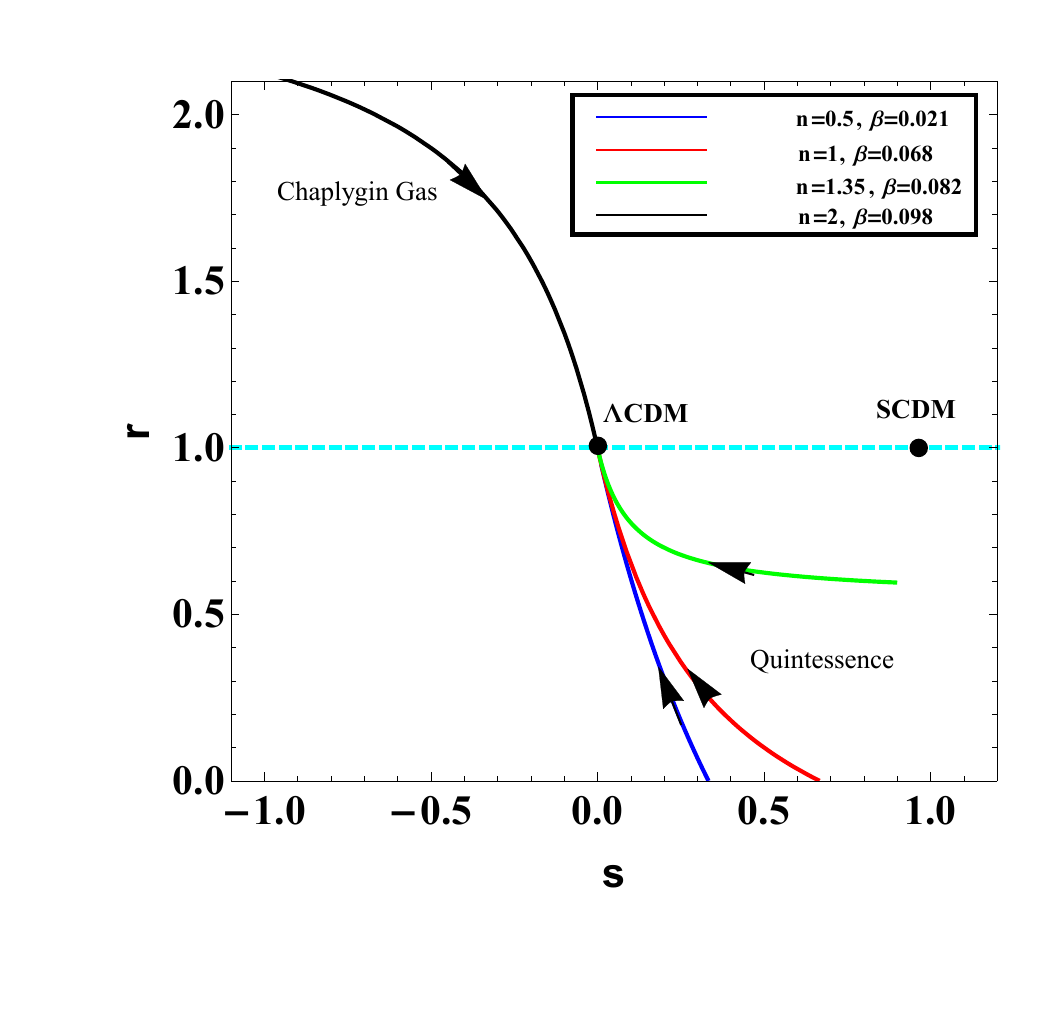}& %
\includegraphics[width=2.5 in, height=2.5 in]{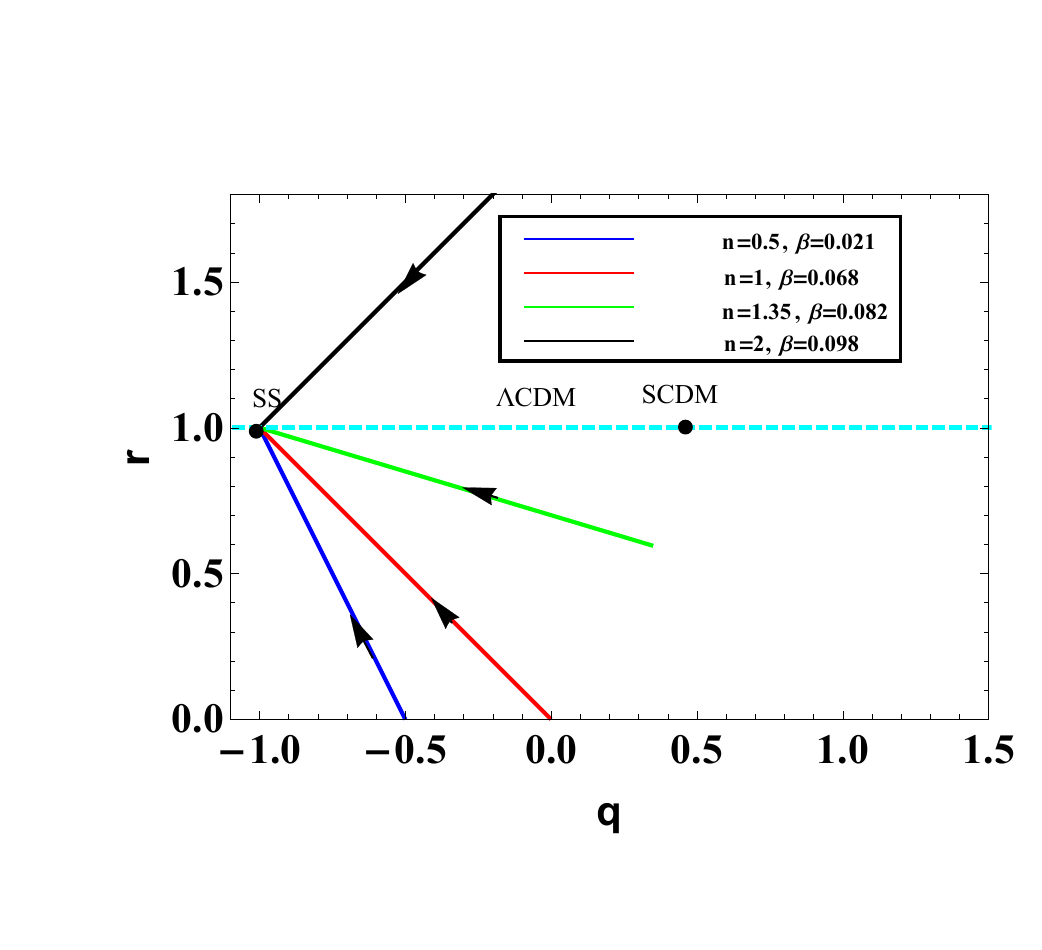}\\
\mbox (a) & \mbox(b)
\end{array}%
$%
\end{center}
\caption{\scriptsize The behavior of $ s-r $ and $ q-r $ trajectories for different values of $ n $ and $ \beta $.}
\end{figure}

Fig. 7(a) highlight the evolution of four trajectories with time for $ n $ and $ \beta $ in $ s-r $ plane. Each trajectory for  $ n=1, \beta=0.068 $;  $ n=1.35, \beta=0.082 $ and $ n=2, \beta=0.098 $ exhibit the  same pattern as all begin in the region $r<1$ and $s>0$ evolving with time, approaches to $\Lambda$CDM model \textit{i.e.} the point $r=1, s=0$ . The time evolution of the trajectory corresponding to $n=0.5, \beta=0.021$ starts from the region $r>1$ and $s<0$ and eventually approaches to $\Lambda$CDM model. From the plot 7(a), we observe that all the trajectories deviate from SCDM which is resemble to matter dominated universe, exhibit different dark energy candidates as Chaplygin gas for $n=0.5, \beta=0.021$, quintessence for $n=1, \beta=0.068$; $n=1.35, \beta=0.082$ and $n=2, \beta=0.098$, $ \Lambda $CDM for $r=1,s=0$ and SCDM for $r=1,s=1$. Thus the various DE scenarios can be observed by these evolutionary trajectories which are the remarkable features of statefinder diagnostic.\\

Fig. 7(b) states the evolution of the four trajectories with time for $ n $ and $ \beta $ in $ q-r $ plane. Each trajectory begins in the neighbourhood of $SCDM$ at the time of evolution of the Universe without passing through $\Lambda$CDM and $SCDM$ converge to $ SS $, the steady state model of the Universe. The downward pattern of a trajectory corresponding to $ n=0.5 $, $ \beta=0.021 $ and upward trend of the trajectories corresponding to $ n=2 $, $ \beta=0.098 $; $ n=1 $, $ \beta=0.068 $ and $ n=1.35 $, $ \beta=0.082 $ converge to the point $ r=1 $, $ q=-1 $ denoted by SS \textit{i.e.} the steady state model of the Universe which suggest the steady state behavior of dark energy model in late times.\\ 

\section{Observational constraints on the model parameters}

\qquad An impressive feature of astronomy is associated with its recent progress in observational cosmology. Study of the origin, evolution, structure formation, properties of dark matter and dark energy in the Universe with the help of cosmic instruments and ray detectors is called observational cosmology. There are several types of observational data available today for different measurements in the field of cosmology. Some of them are Sloan Digital Sky Survey ($ SDSS $) which provide the map of the galaxy  distribution and encode the current fluctuations in the Universe,$ CMBR $ that serves as  the evidence of the big bang theory, Quasi Stellar Radio Sources ($ QUASARS $) which are considered as the most metal thing in the Universe and extract the matter between observer and quasars, Baryon Acoustic Oscillations ($ BAO $) that measures the large scale structures in the Universe in order to understand dark energy better, observations from type Ia Supernova are the tools for measuring the cosmic distances usually known as standard candles. In the subsequent sections, we have presented a statistical analysis by using some observational datasets of $ SNeIa $, $H(z)$ and $ BAO $ to constrain Hubble parameter $H_0$ and model parameter $n$ involved in our model. To constraint model parameter $n$, we restrict the inverse hyperbolic of sine series and hyperbolic of cotangent series in Eq. (\ref{41}) upto first term and then integrate the approximate series to calculate the Chi-square value \textit{i.e.} $ \chi^2_{min} $ using each observational data set.\\

\begin{figure}[tbph]
\begin{center}
$%
\begin{array}{c@{\hspace{.1in}}c}
\includegraphics[width=2.5 in, height=2.3 in]{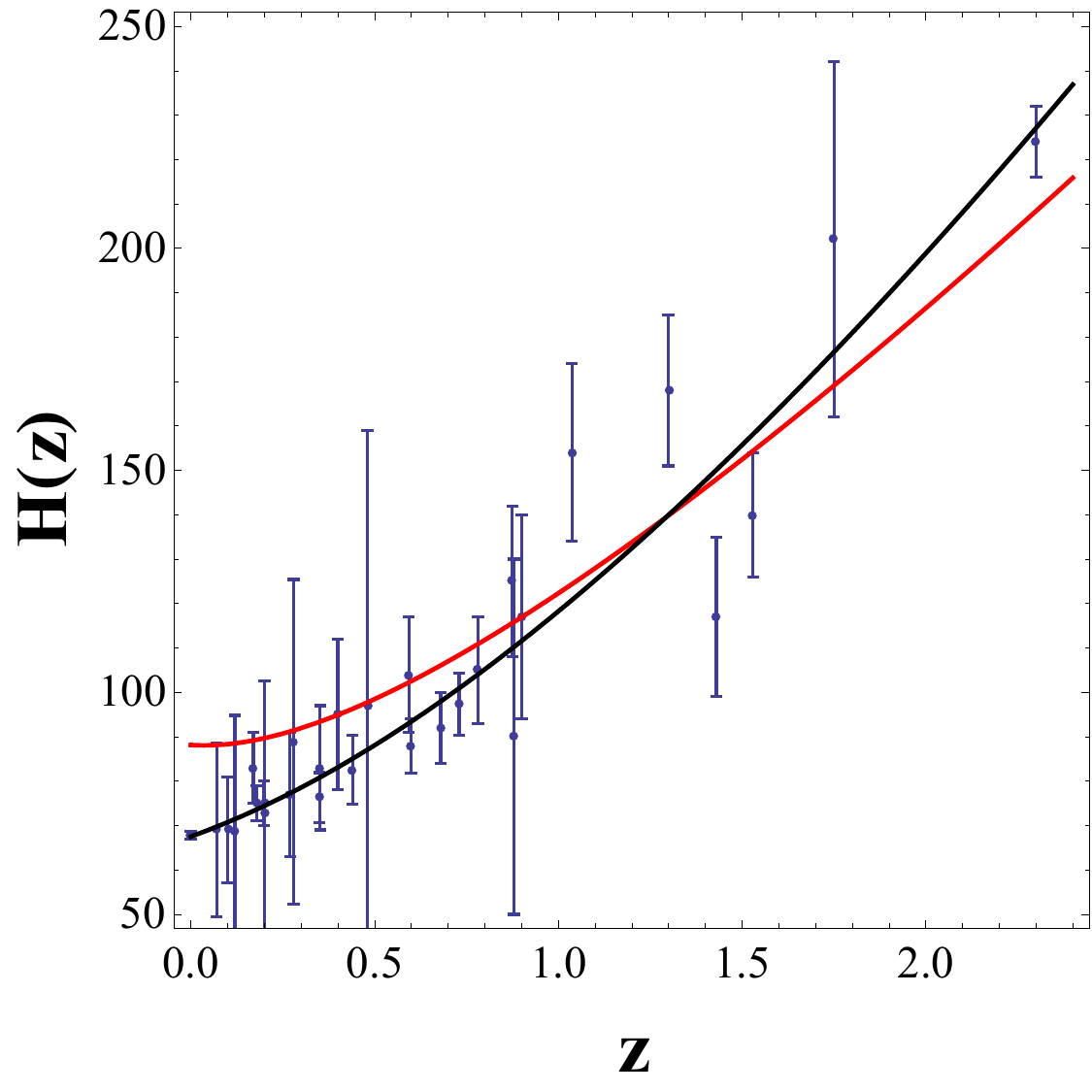} & %
\includegraphics[width=2.5 in, height=2.3 in]{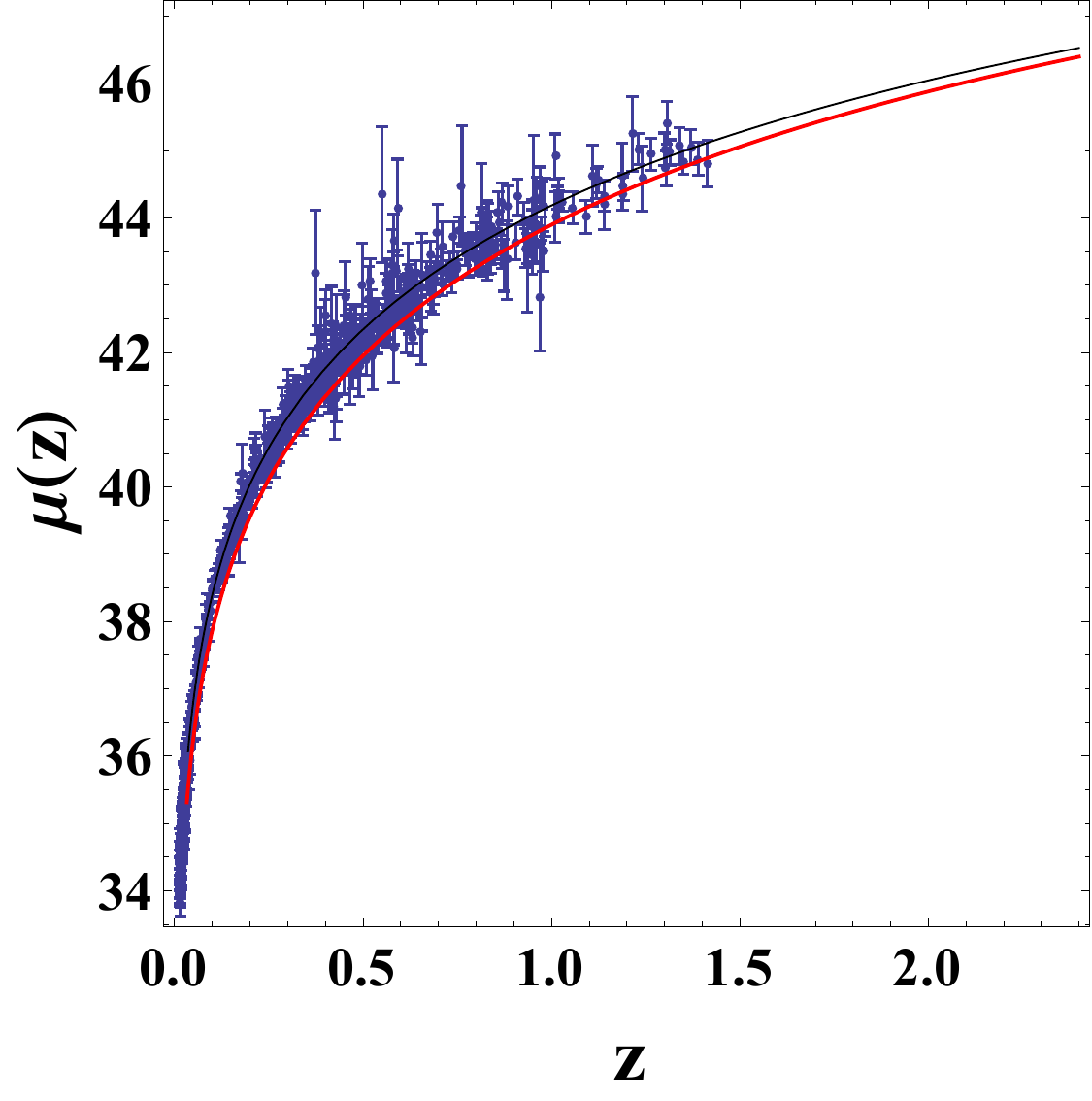} \\ 
\mbox (a) & \mbox (b)%
\end{array}%
$%
\end{center}
\caption{\scriptsize Figures (a) and (b) are error bar plots comparing our model with standard $\Lambda$CDM model using $ H(z)$ and $ SNeIa$ datasets respectively. Black line signifies $ \Lambda CDM $ model and red line displays our model in both figures (a) and (b).}
\end{figure}

\subsection{Hubble observation H(z)}
The Hubble parameter $H$ can be observed in terms of some physical observable quantity such as length, time and redshift $z$. In terms of redshift $z$, $H$ reads 
\begin{equation}\label{hzfunc}
H(z) = -\frac{1}{(1+z)} \frac{dz}{dt}.
\end{equation}
Here, the best fit value of model parameter $n$ is obtained by using $ 28 $ points of $ H(z) $ datasets \cite{Hz} in the range $ (0.1,2.5) $. We take the recent value of Hubble constant $H_{0}=67.8$ $ Km/s/Mpc $ from Planck $2014$ results \cite{Hz-Plank} to compare our model with standard $\Lambda$CDM. The constrains on model parameter $n$ can be obtained by minimizing the Chi-square value \textit{i.e.} $\chi^2_{min} $, which is identical as the maximum likelihood analysis and is expressed as


\begin{equation}\label{oc1}
\chi _{OH}^{2}=\sum\limits_{i=1}^{28}\frac{[H^{obs}(z_{i})-H^{th}(z_{i})]^{2}}{\sigma {(z_{i})}^{2}},
\end{equation}
where, $ OH$ indicates the observational Hubble dataset. $ H^{obs} $ and $ H^{th} $ serve the observed and theoretical value of $H $. $\sigma{(z_{i})}$ denotes the standard error in the measured value of $H$. 

\subsection{ Type Ia Supernova}
Here, we fit the present model with latest union $2.1$ compilation observational dataset of $580$ point \cite{SNeIa} and compare the results with $\Lambda$CDM.  
\begin{equation}\label{oc2}
\chi _{OSN}^{2}(\mu_0)=\sum\limits_{i=1}^{580}\frac{[\mu_{th}(\mu_0,z_{i})-\mu_{obs}(z_{i})]^{2}}{\sigma _{\mu(z_{i})}^{2}},
\end{equation}
where $OSN $ denotes the observational $ SNeIa $ dataset. $\sigma_{\mu(z_{i})}$, $\mu_{obs}$ and $\mu_{th}$ denote the standard error in the measurement of $\mu(z)$, the observed and theoretical distance modulus of the model respectively. The distance model $\mu(z)$ is given by 
\begin{equation}\label{oc3}
\mu(z)= m-m' = 5Log D_l(z)+\mu_{0},
\end{equation}
where $D_l(z) $ and $\mu_0$ are the luminosity distance and nuisance parameter respectively. $m$ and $m'$ represent the apparent and absolute magnitudes of standard candle respectively.   \\

The left panel of Fig. 8 is not good fitted so much but right panel is better fitted when we compare our model with $\Lambda$CDM using $H(z)$ and SNeIa data respectively.


\begin{figure}[tbph]
\begin{center}
$%
\begin{array}{c@{\hspace{0.1 in}}cc}
\includegraphics[width=2.2 in, height=2.0 in]{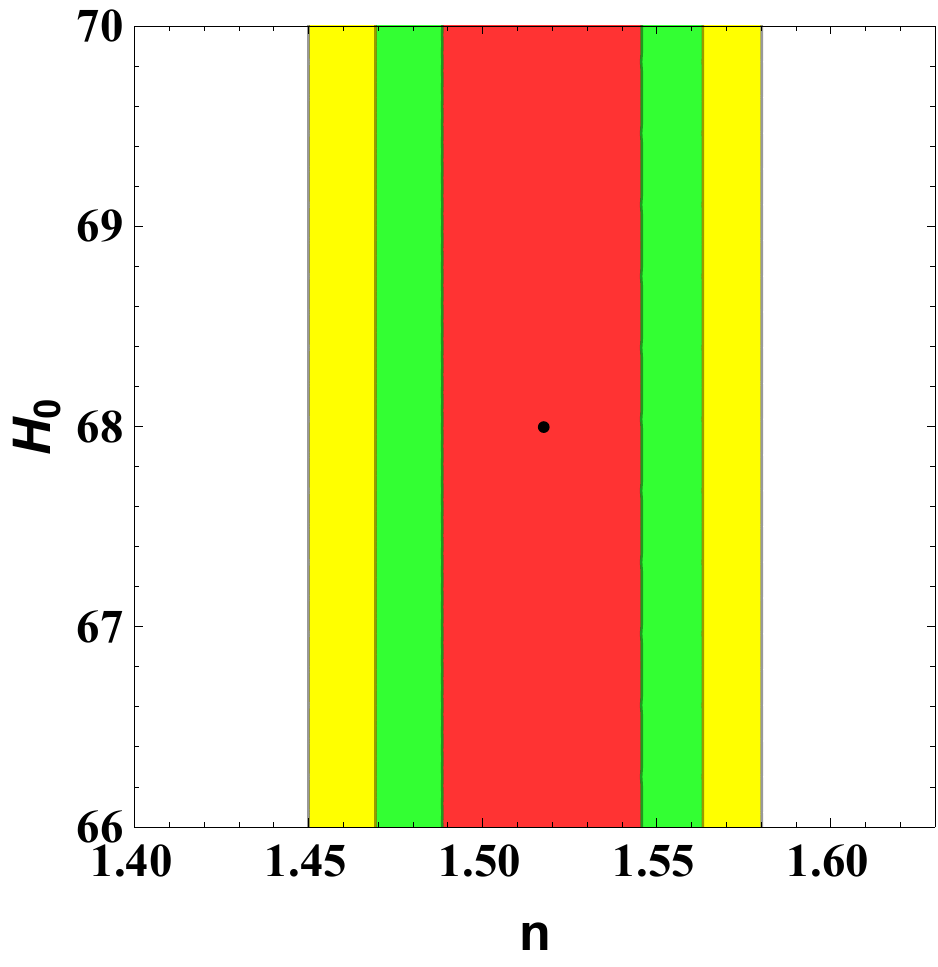} & 
\includegraphics[width=2.2 in, height=2.0 in]{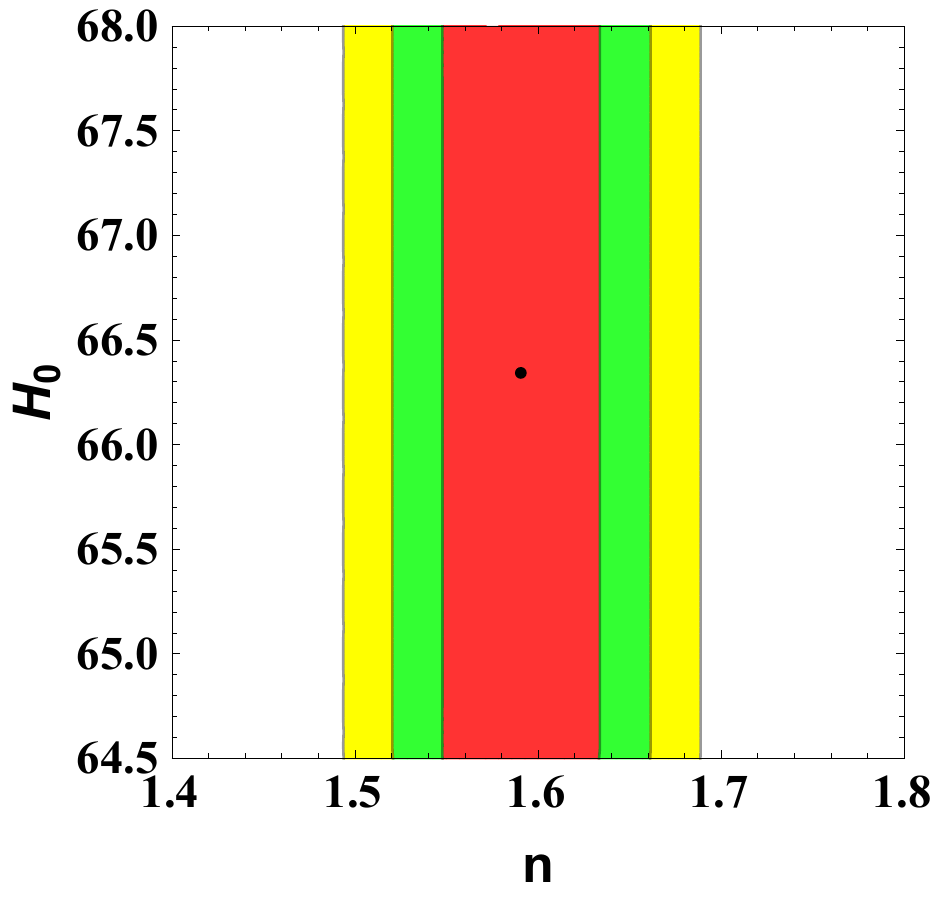} &
\includegraphics[width=2.2 in, height=2.0 in]{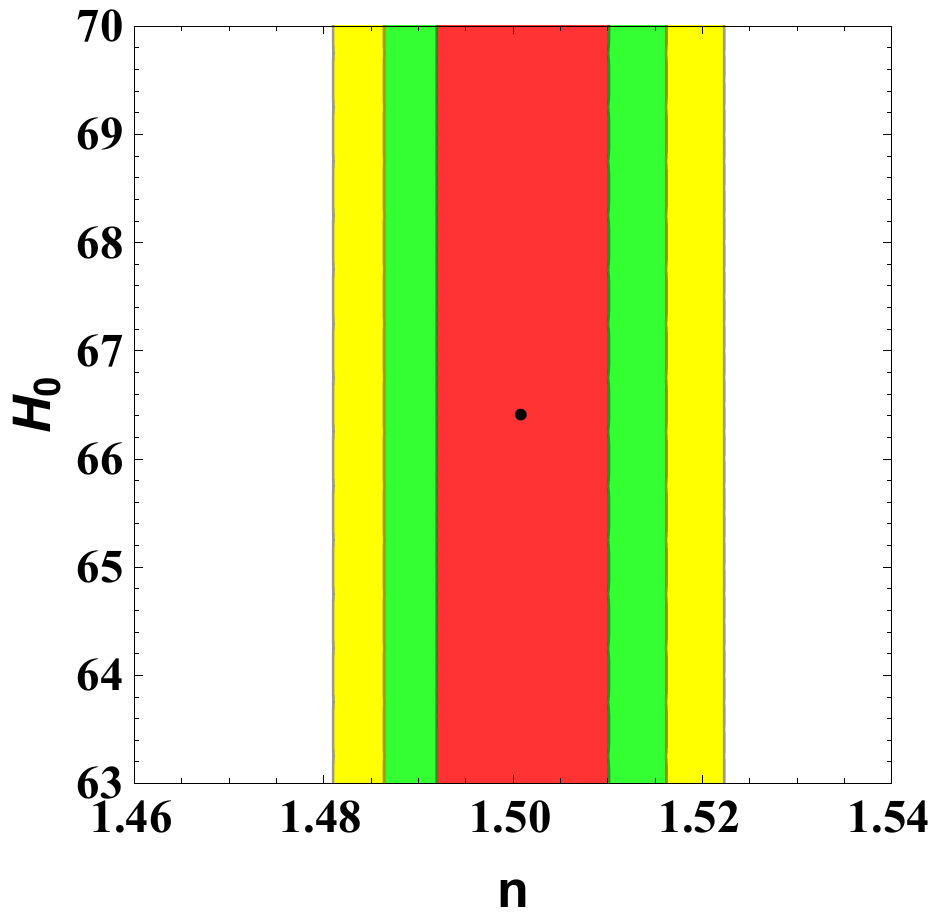} \\ 
\mbox (a) & \mbox (b) & \mbox(c)
\end{array}%
$%
\end{center}
\caption{\scriptsize Plots (a), (b) and (c) represent the contours in $( n $-$ H_{0} )$ plane from the statistical analysis of  $ H(z)$, $ SNeIa $ and $ BAO $ datasets respectively with 1-$ \sigma $ error (red shaded region ), 2-$ \sigma $ error (light green shaded region shows) and 3-$\sigma $ error (light yellow shaded region). Black dot signifies the the well fitted constrained value of model parameter $ n $ and the values of $ H_{0} $ for each observation.}
\end{figure}

\subsection{ Baryon Acoustic Oscillations}
The $BAO$ measures the structures of the universe on a large scale taking the statistical property of matter as sound waves which is useful to study DE in a better way. Here, we consider a sample from various surveys, namely $6dF$ Galaxy survey \cite{6df}, $SDSS$ galaxy sample \cite{padn}, results from $ WiggleZ $ survey \cite{wig} and $BOSS CMASS $ \cite{boss}.\\

\begin{figure}[tbph]
\begin{center}
$
\begin{array}{c@{\hspace{0.1in}}c}
\includegraphics[width=2.5 in, height=2.3 in]{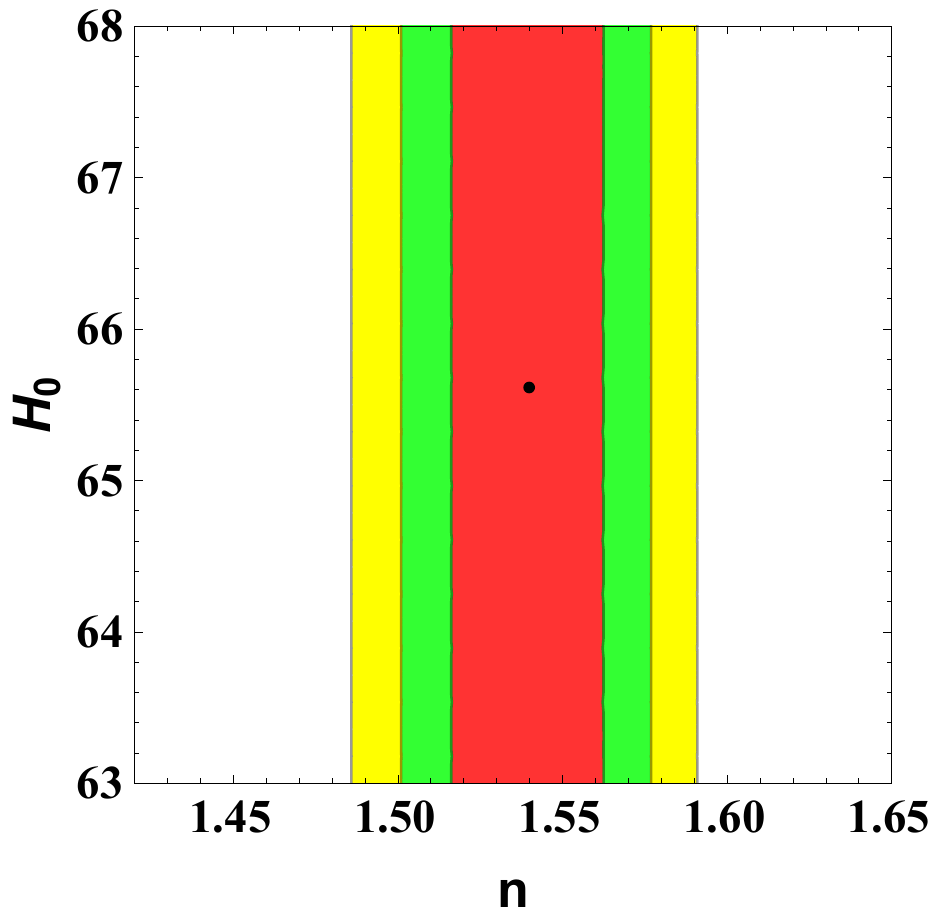} & 
\includegraphics[width=2.5 in, height=2.3 in]{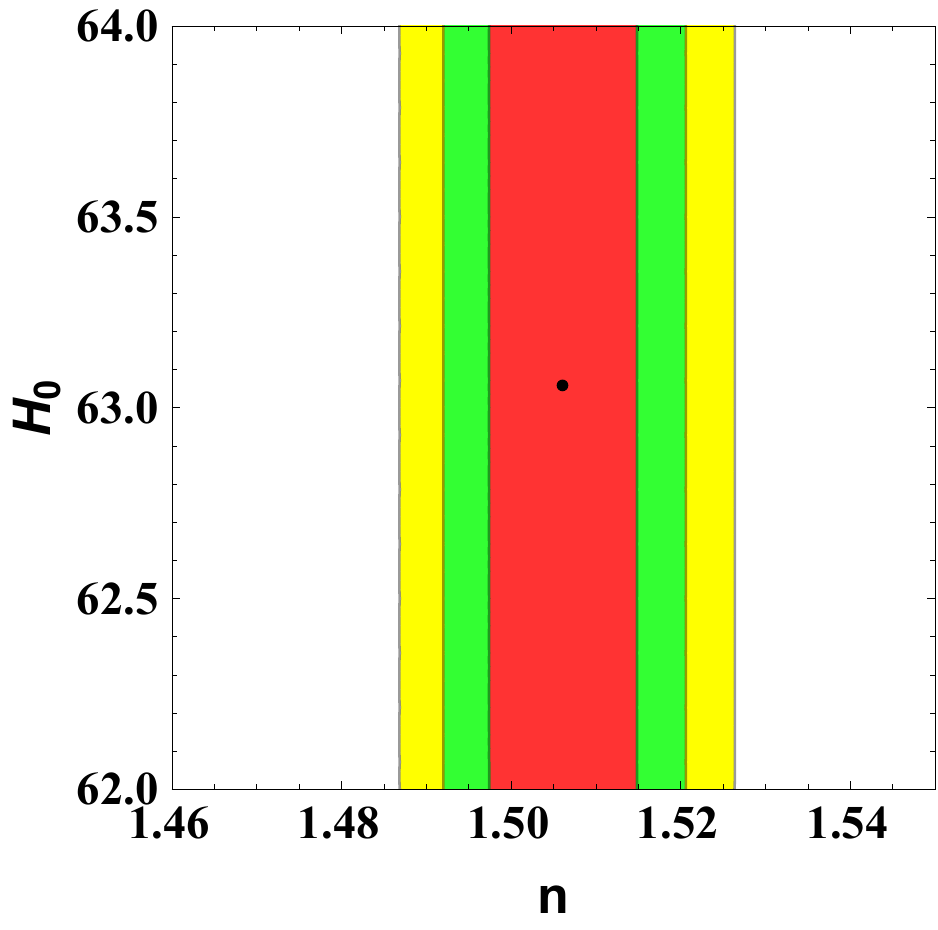} \\ 
\mbox (a) & \mbox (b) 
\end{array}
$
\end{center}
\caption{\scriptsize Plots (a), (b) and (c) represent the contours in $( n $-$ H_{0} )$ plane from the statistical analysis of   $ H(z)$ + $ SNeIa $  and $ H(z)$ + $ SNeIa $  + $ BAO $ respectively with 1-$ \sigma $ error (red shaded region ), 2-$ \sigma $ error (light green shaded region shows) and 3-$\sigma $ error (light yellow shaded region). Black dot signifies the well fitted constrained value of model parameter $ n $ and the values of $ H_{0} $ for each observation.}
\end{figure}

We define $d_z$ (Distance redshift ratio) for $BAO$ measurement 
\begin{equation}\label{drr}
d_z=\frac{r_s(z_*)}{D_v(z)},
\end{equation}

where $ z_* $ denotes the redshift at the time of photons decoupling redshift which is $ z_*=1090 $ given by Planck 2015 results \cite{adep}. Also $ r_s(z_*) $ is co-moving sound horizon during decoupling of photons \cite{waga}. Further the dilation scale $D_v(z)= \big[\frac{d^2_A(z) z}{H(z)}\big]^{\frac{1}{3}}$, where $d_A(z)$ indicates the angular diameter distance.\\

The Chi square value for $ BAO $ measurement $(\chi _{BAO}^{2})$  is given by \cite{gio}
\begin{equation}\label{OBAO}
\chi _{BAO}^{2}= Y^{T}C^{-1}Y,
\end{equation}
where $ Y $ is a matrix given by
 \[
   Y=
  \left[ {\begin{array}{cc}
   \frac{d_A(z_*)}{D_v(0.106)}-30.84  \\
   \frac{d_A(z_*)}{D_v(0.35)}-10.33 \\
   \frac{d_A(z_*)}{D_v(0.57)}-6.72 \\
   \frac{d_A(z_*)}{D_v(0.44)}-8.41 \\
   \frac{d_A(z_*)}{D_v(0.6)}-6.66 \\
   \frac{d_A(z_*)}{D_v(0.73)}-5.43 \\
  \end{array} } \right]
\]
The inverse covariance matrix $C^{-1}$ \cite{gio} by adopting the correlation coefficients existed in \cite{hing} becomes
\[
  C^{-1}=
  \left[ {\begin{array}{cccccc}
   0.52552 & -0.03548 & -0.07733 & -0.00167 & -0.00532 & -0.00590\\
   -0.03548 & 24.97066 & -1.25461 & -0.02704 & -0.08633 & -0.09579\\
   -0.07733 & -1.25461 & 82.92948 & -0.05895 & -0.18819 & -0.20881\\
   -0.00167 & -0.02704 & -0.05895 & 2.91150 & -2.98873 & 1.43206\\
  -0.00532 & -0.08633 & -0.18819 & -2.98873 & 15.96834 & -7.70636\\
  -0.00590 & -0.09579 & -0.20881 & 1.43206 & -7.70636 & 15.28135\\
  \end{array} } \right]
  \]

\begin{table}
\caption{Outcomes of data analysis for our model}
\begin{center}
\label{tabparm}
\begin{tabular}{l c c c r} 
\hline\hline
\\ 
{Data} \,\,\,\,\,  &  \,\,\,\, $ \chi^2_{min} $ \,\,\,\,\,  &  \,\, \, $ H_0 $  \,\,\,\,\,   &  \,\,\,\,\, $  n $ \,\,\,\\ 
\\
\hline 
\\
{$ H(z) $}   \,\,\,\,\,  &  \,\, \,   $ 19.5872 $   \,\,\,\,\,  &  \,\, \,   $ 68.0000 $   \,\,\,\,\,  &  \,\, \,   $ 1.5176 $
\\
\\
{$ SNeIa $}   \,\,\,\,\,  &  \,\, \,    $ 565.5074 $   \,\,\,\,\,  &  \,\, \,   $ 66.3402 $   \,\,\,\,\,  &  \,\, \,   $ 1.5907 $ 
\\
\\
{$ BAO $}   \,\,\,\,\,  &  \,\, \,   $ 2.1622 $   \,\,\,\,\,  &  \,\, \,   $ 66.4138 $   \,\,\,\,\,  &  \,\, \,   $ 1.5009 $ 
\\
\\
{$ H(z) $ + $ SNeIa $ }  \,\,\,\,\,  &  \,\, \,  $ 589.7361 $  \,\,\,\,\,  &  \,\, \,  $ 65.6148 $  \,\,\,\,\,  &  \,\, \,  $ 1.5396 $
\\
\\
{$ H(z)$ + $ SNeIa $ + $ BAO $}   \,\,\,\,\,  &  \,\, \,   $ 597.2995 $   \,\,\,\,\,  &  \,\, \,   $ 63.0579 $   \,\,\,\,\,  &  \,\, \,   $ 1.5060 $
\\
\\ 
\hline\hline  
\end{tabular}    
\end{center}
\end{table}


Figures $9, 10$ represent the contours with error $ 1-\sigma $, $ 2-\sigma $ and  $ 3-\sigma $ in $(n $-$ H_{0})$ plane. The well fitted constrained value of the model parameter $ n =1.5176 $ and Hubble constant $ H_{0}=68.0000 $ $ Km/s/Mpc$ with $ 1\,\sigma $, $ 2\,\sigma $ and  $ 3\,\sigma $ errors are found in the ranges $ 1.488<n<1.5461 $, $ 1.469<n<1.563 $ and $ 1.449<n<1.58 $ respectively according to the Hubble dataset $H(z)$. The well fitted constrained value of the model parameter $ n =1.5907 $ and Hubble constant $ H_{0}=66.3402 $ $ Km/s/Mpc$ with $ 1\,\sigma $, $ 2\,\sigma $ and  $ 3\,\sigma $ errors are found in the ranges $ 1.547<n<1.634 $, $ 1.521<n<1.662 $ and $ 1.493<n<1.69 $ respectively according to the $ SNeIa $ (union 2.1 compilation data set). The well fitted constrained value of the model parameter $ n =1.5009 $ and Hubble constant $ H_{0}=66.4138 $ $ Km/s/Mpc$ with $ 1\,\sigma $, $ 2\,\sigma $ and  $ 3\,\sigma $ errors are found in the ranges $ 1.492<n<1.510 $, $ 1.486<n<1.516 $ and $ 1.481<n<1.522 $ respectively according to the $ BAO $. The well fitted constrained value of the model parameter $ n =1.5396 $ and Hubble constant $ H_{0}=65.6148 $ $ Km/s/Mpc$ with $ 1\,\sigma $, $ 2\,\sigma $ and  $ 3\,\sigma $ errors are found in the ranges $ 1.516<n<1.563 $, $ 1.501<n<1.577 $ and $ 1.486<n<1.591 $ respectively according to the combined dataset $ H(z)$ + $ SNeIa $. The well fitted constrained value of the model parameter $ n =1.5060$ and Hubble constant $ H_{0}=63.0579 $ $ Km/s/Mpc$ with $ 1\,\sigma $, $ 2\,\sigma $ and  $ 3\,\sigma $ errors are found in the ranges $ 1.497<n<1.515 $, $ 1.492<n<1.521 $ and $ 1.487<n<1.526 $ respectively according to the combined dataset $ H(z)$ + $ SNeIa $ + $ BAO $ (see Table IV). \\

\section{ Some other cosmological tests}

\subsection{ Lookback time}

\qquad The time interval between the detection of light on the Earth and the emission from the source is called Lookback time denoted by $ t_L $. Therefore, the total time $t_L$ elapsed between the light ray which emits from a galaxy at time $ t_z $ at particular redshift $ z $ and reaches us at a time $ t_0 $ at redshift $ z=0 $ is given by \\
\begin{equation}\label{56}
t_L= t_0-t_z=\int_{a}^{a_0} \frac{dt}{\dot{a}}.
\end{equation}
From (\ref{40}), we have
\begin{equation} \label{57}
t_L=t_0-\frac{sinh^{-1}\sqrt{\frac{n-(1+q_0)}{(z+1)^{2n}(q_0+1)}}}{\beta}.
\end{equation}

\subsection{ Proper distance}

\qquad The proper distance is the distance travelled  by the photons from the source to us which is also known as \textit{Instantaneous distance}. Proper distance is denoted by $ d(z) $ and it can be calculated as $ d(z)=a_0 r(z) $, where $ r(z) $ is the radial distance which can be obtained by
\begin{equation}\label{58}
r(z)=\int_t^{t_0} \frac{dt}{a(t)}.
\end{equation}\\
Using Eq. (\ref{40}), the proper distance $ d(z) $ of the model is obtained by  
\begin{equation}\label{59}
d(z)=\frac{1}{\beta}\Bigg[-cosh(\beta t)\,S_1\, sinh(\beta t)^{-\frac{1+n}{n}}(-sin h(\beta t)^2)^{\frac{1+n}{2n}}+cosh(\beta t_0)\,S_2\, sinh(\beta t_0)^{-\frac{1+n}{n}}(-sin h(\beta t_0)^2)^{\frac{1+n}{2n}}\Bigg],
\end{equation}
where $S_1, S_2$ are hypergeometric functions defined as
$$S_1=Hypergeometric\,\,2F1[\frac{1}{2},\frac{1}{2}(1+\frac{1}{n}),\frac{3}{2}, cosh(\beta t)^2]$$ and $$S_2=Hypergeometric\,\,2F1[\frac{1}{2},\frac{1}{2}(1+\frac{1}{n}),\frac{3}{2}, cosh(\beta t_0)^2].$$

\begin{figure}[tbph]
\begin{center}
$%
\begin{array}{c@{\hspace{.1in}}cc}
\includegraphics[width=2.5 in, height=2.0 in]{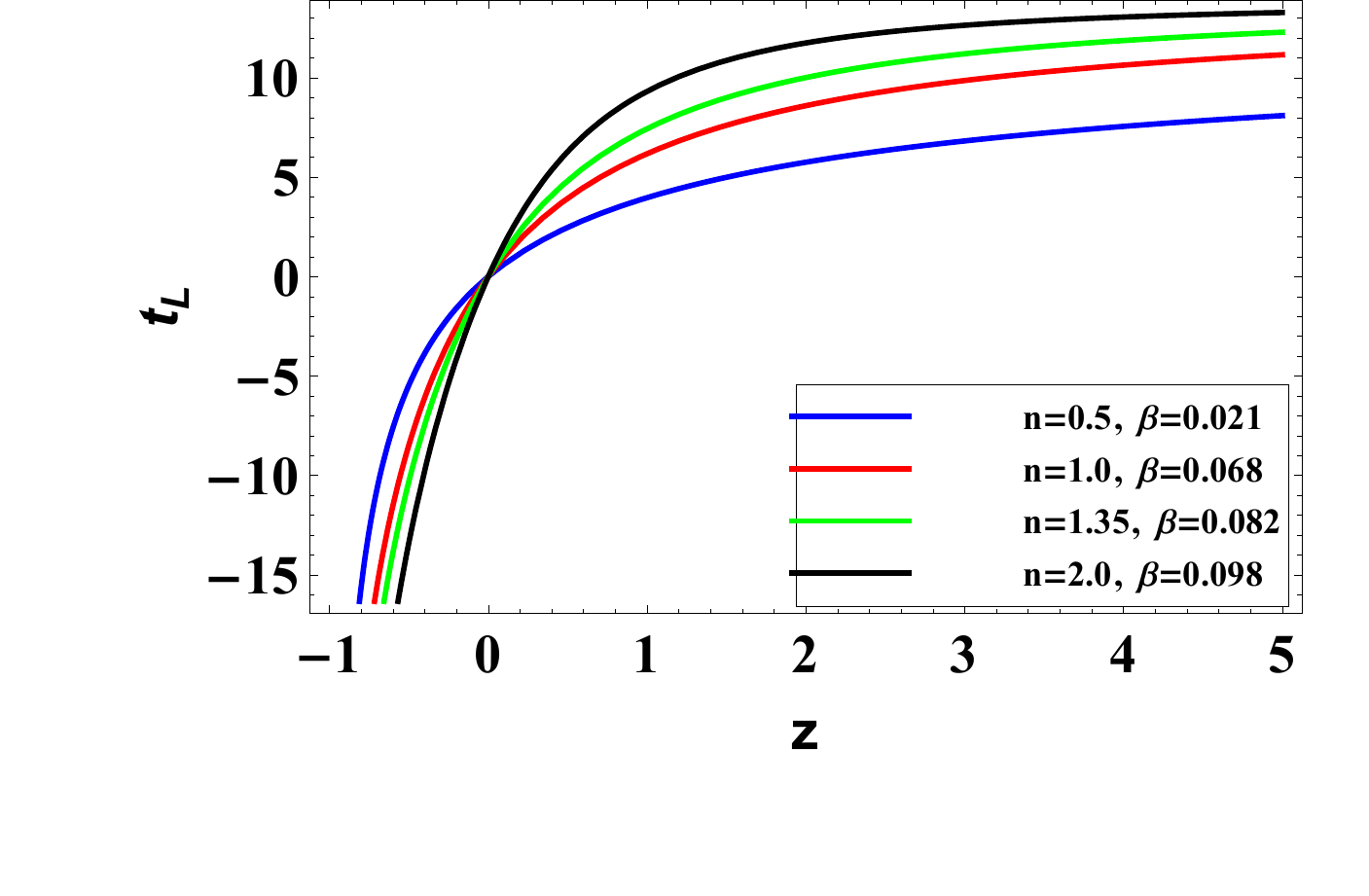} & %
\includegraphics[width=2.5 in, height=2.0 in]{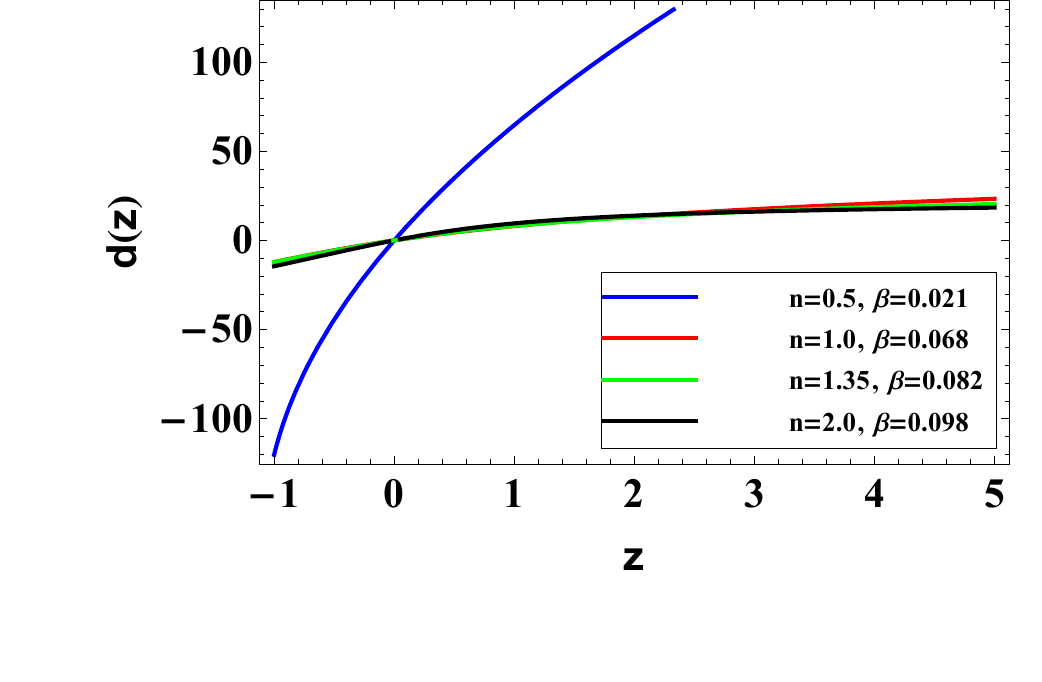} &  \\
\mbox (a) & \mbox (b) &
\end{array}%
$%
\end{center}
\caption{\scriptsize The plots of look back time $t_{L}$ and proper distance $d(z)$ Vs. redshift $z$.}
\end{figure}

\subsection{ Angular diameter}

\qquad If the size of a source is known, then its angular width which is its distance from the observer is given by
\begin{equation}\label{60}
d_A=\frac{\theta}{l},
\end{equation}
where $\theta$ is the angular size of the source and $l$ is the size of the source. In terms of redshift $z$, the angular diameter is given by
\begin{equation}\label{61}
d_A= \frac{d(z)}{1+z}.
\end{equation}

\subsection{ Luminosity distance}

\qquad Suppose an object is located at a distance $r$ with absolute luminosity $L$ and observed luminosity $l$ of the source then we have 
\begin{equation}\label{62}
L=\frac{l}{4\pi r^2}.
\end{equation}\\
and the luminosity distance $ d_l $ is defined as\\
\begin{equation}\label{63}
d_l=\left(\frac{L}{4\pi l}\right)^\frac{1}{2}.
\end{equation}\\
In terms of $z$, the luminosity distance $ d_l $ is given by
\begin{equation}\label{64}
d_l=(1+z) d(z).
\end{equation}
\begin{figure}[tbph]
\begin{center}
$%
\begin{array}{c@{\hspace{.1in}}cc}
\includegraphics[width=2.5 in, height=2.0 in]{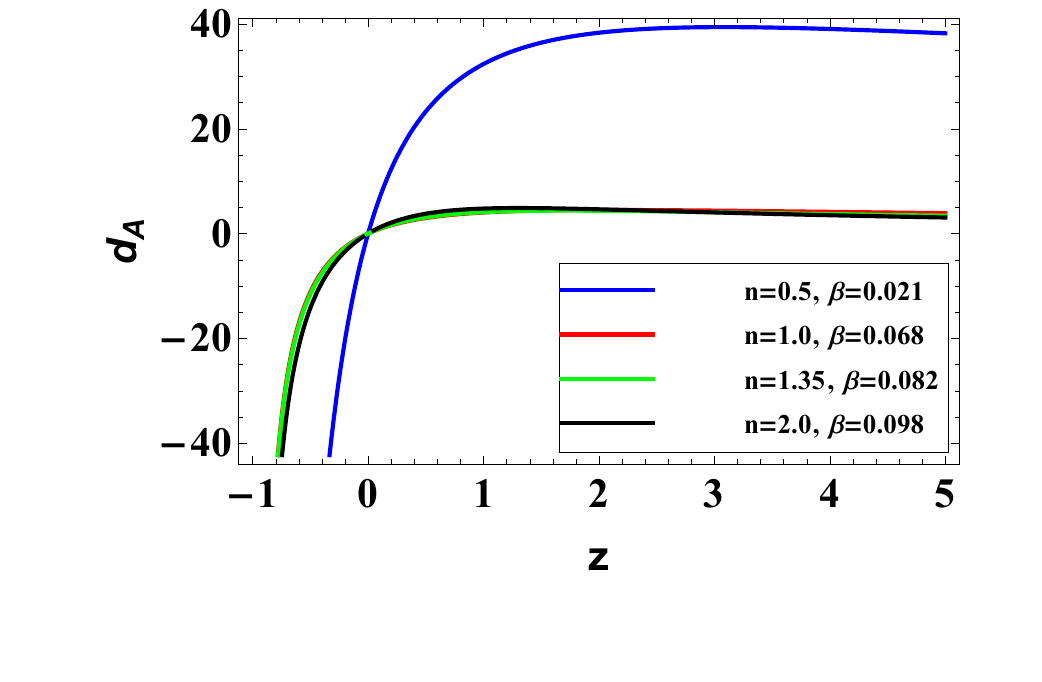} & %
\includegraphics[width=2.5 in, height=2.0 in]{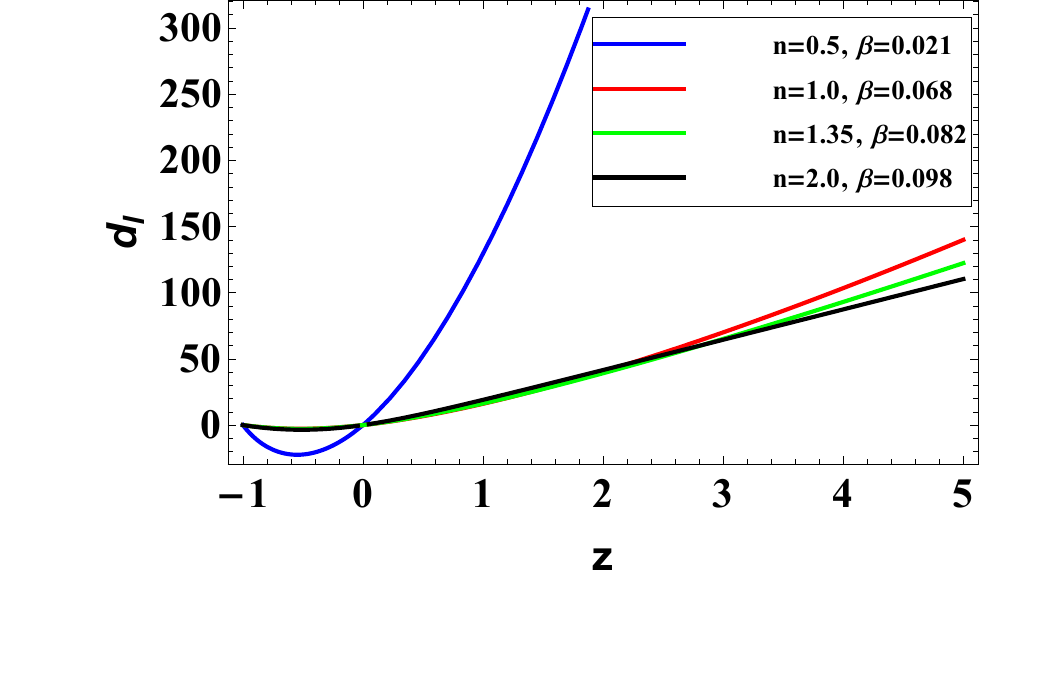} &  \\
\mbox (a) & \mbox (b) &
\end{array}%
$%
\end{center}
\caption{\scriptsize The plots of angular diameter distance $d_{A}$ and luminosity distance $d_l$\ Vs. redshift $z$.}
\end{figure}

\section{ Discussions and Conclusions}

\qquad In present paper, we have examined the cosmological model in the background of FLRW metric under the theory of $ f(R,T) $ gravity with the form $ f(R,T)=f(R)+2f(T) $. The functional form of $ f(R) $ is considered in a sense that it provides quadratic geometric corrections in GR and it is accepted as the most compatible model with the latest observations. The variable $ T $ is the trace of EMT, which introduce exotic imperfect fluids and quantum effects. Also the functional form of $ T $ explains the non minimal matter-geometry coupling discussed in \cite{rap}. Consequently the effect of complete form of $ f(R,T) $ function on the general action produces an explicit set of field equations. Therefore, the behavior of the Universe depends on the field equations and based on the associated informations and the observations using various statistical data can be interpreted as follows.
\begin{itemize}
\item The geometrical parametrization of scale factor $ a(t) $ has been chosen in such a way that it leads to two types of cosmic models depending upon various the range of $ n $. The time dependent DP shows different regimes of the Universe as (i) the model shows eternal acceleration when $0<n<1$ and (ii) the model transits from early deceleration to late time acceleration of the Universe which can be seen in Fig 1(c). Hence the phase transition occurs when $ n\geq 1 $. For $ n=1.35 $, $ q=0 $ at redshift transition values $ z_{tr}=0.883752 $ and for $ n=2 $, $ q=0 $ at redshift transition values $ z_{tr}=0.352666 $, which are best fitted redshift transition values according as the observations \cite{abd, zha} respectively. Also we inspect that the model exhibits point type singularity at time $ t=0 $ (see Table I).

\item Using the parametrization of $a(t)$, we have obtained the solution of the field equations (\ref{19})-(\ref{20}) \textit{i.e.} the value of energy density $\rho$ and isotropic pressure $p$ of the matter filled in the Universe. The plots of $\rho$ and $p$ can be seen  in Fig 2. Fig. 2(a) represents the behavior of $\rho$ corresponding to $ n = 0.5,\, 1,\, 1.35,\, 2 $. Initially at the point of singularity $t=0$, the Universe starts with very high energy density  $\rho\to \infty$ and later on falls off as time unfolds $\rho \to 0$ in late times $t\to \infty$. Fig. 2(b) highlights the profile of pressure for all the values of $n$ as mentioned earlier. The isotropic pressure $p \to -\infty$ as $t\to 0$, remains negative throughout the evolution and approaches to negative constant value in late times for $n=0.5$, which indicates the eternal cosmic accelerated expansion. For $ n=1 $, $ p $ varies from initially positive value in a very short interval of times to a negative value in late times and sequentially approaches to a negative constant. For $ n= 1.35, 2 $, the isotropic pressure $ p $ decreases monotonically with time and attains a negative value in late times. The initially small positive value of pressure resembles the decelerating expansion of the Universe while according to standard cosmology, the negative pressure is indeed responsible for cosmic acceleration. The model is rational with formation of structure in the Universe for the specified values of $ n = 1,\, 1.35,\, 2 $.

\item The result (\ref{434}) depends on the coupling constant of the matter part $\lambda$. From (\ref{434}) we see that matter behaves like the DE in late times and the value (\ref{434}) is observationally accepted for the matter density and pressure in late times provided $\lambda>-2\pi$. More precisely, the limit value of the matter density in the early times is the necessary condition that our model satisfies the weak energy condition (WEC), provided the constraints in the Table III must hold.

\item From Eqs. (\ref{434-1}) and (\ref{434-2}), we observe that the EoS parameter (\ref{43}) tends to a constant value in the early times and it shows the DE in late times. From Eq. (\ref{434-1}) we find two different cosmological transitions in our model: (i) a transition from a stiff matter dominants at the early stage to a DE dominants in late times when $ n=3/2 $ , and (ii) a transition from a pressure less matter dominated era with $\omega^{p}_{i}=0$ in the early eras to a DE dominants in late times when $ n=\frac{3 (\lambda +4 \pi )}{2 (3 \lambda +8 \pi )} $. Therefore, it is interesting that the final state of cosmological evolution depends on $ \lambda $ only. In Fig. 3., for a fix value of $\alpha=0.5$, and $\lambda=2$, $\omega \in $ quintessence region for high redshift $z$ and as time unfolds, $\omega \to -1$ in infinite future (\textit{i.e.} $ z\to -1$) when $n$ equals to the expression (\ref{434-3}), which is presently consistent with the observations of temperature fluctuation in $ CMBR $ \cite{hin}. The current value of the $\omega$ for the mentioned types of evolution era are obtained in Eqs.  (\ref{434-4}) and  (\ref{434-5}) when $ n=3/2, \,\frac{3 (\lambda +4 \pi )}{2 (3 \lambda +8 \pi )} $respectively. Therefore, the astronomical data can constrain the coupling constant. From Fig. 3(b), the EoS parameters $ \omega\simeq-0.509 $ and  $ \omega\simeq-0.2629 $ at present epoch $z=0$ for $ n=3/2,\, \frac{3 (\lambda +4 \pi )}{2 (3 \lambda +8 \pi )} $ respectively, which is in good agreement with the observations \cite{abd}.

\item After inspecting the EoS parameter $\omega$, it can easily be observed that there is some exotic matter field present in the Universe which is capable of being produced some repulsive pressure. The matter field has negligible $ KE $ as compare to $ PE $ in the Universe which makes $\omega$ negative. The behavior of the quintessence scalar field correspondence has been examined in subsection IV A. In Fig. 4, we observe that $ V(\phi) $ is present in the interval $ -1<\phi<0 $ and $ V(\phi)\simeq0 $ at $ \phi\simeq0 $. Therefore, we can predict that the $ \phi $ is the only source of DE with $ V(\phi) $. Thus we conclude that our model is an accelerating dark energy model.

\item In subsection IV B, we have discussed the physical acceptability of the model by analysing EC. The derived model for $ n=0.5,\,1,\,1.35,\,2 $ satisfy NEC, WEC and DEC for a wide accepted range of $\lambda$. On the other hand, SEC fails to satisfy in the late stages of the Universe, which produces a repulsive force and make the Universe to get jerk. Also the validation of SEC in the early stage leads  to the early decelerated phase of the Universe (See Fig. 5).

\item In section V, we have examined the validity of our model through the various observational tests as: (i) Analysis of Jerk parameter, (ii) Om diagnostic (iii) Analysis of velocity of sound.\\

Fig 6(a) enacts the evolution of jerk parameter $ j $ \textit{w.r.t.} redshift $ z $ for the given parametrization of DP for the values of $ n=0.5,\,1,\,1.35,\,2 $. Jerk parameter has attained positive values for any redshift $z$ and $j \to 1$ represents $ \Lambda CDM $ in the future at $z\to 1$. At present $z=0$, the value of $j$ for $n=0.5,1,1.35$ is clearly less than $1$ and for $n=2$, the value of jerk parameter, which is greater than $1$, favours a dark energy model different from $ \Lambda CDM $.\\

Fig 6(b) expresses the behavior of different dark energy model for different values of $n$. The model exhibits quintessence-like behavior when $ n=0.5,\,1,\,1.35 $ and the phantom-like behavior when $n=2$. Further by considering different values of $\lambda$, we have different stability scenario of model. Model agrees on $ C_s^2 \leq 1 $ throughout the evolution with time $ t $ for all the cases $ n=0.5,\,1,\,1.35,\,2 $ when $\lambda$ assumes value in the range $-17.5\leq \lambda \leq -12.57$ (See Fig. 6(c)). Otherwise, $0<C_s^2<1$ does not hold in other interval. \textit{i.e.} the model is partially stable.

\item  From the Fig. 7(a), we observe that all the trajectories approach to $ \Lambda CDM $ and deviate from $ SCDM $ which is resemble to matter dominated universe. The trajectories exhibit different dark energy candidates as Chaplygin gas for $ n=0.5, \beta=0.021 $, quintessence for $ n=1, \beta=0.068 $; $ n=1.35, \beta=0.082 $ and $ n=2, \beta=0.098 $, $ \Lambda CDM $ for $r=1,s=0$ and $ SCDM $ for $ r=1,s=1 $. Thus the various DE scenarios can be observed by these evolutionary trajectories, which are the remarkable features of statefinder diagnostic. Figure 7(b) states the all the trajectories begin in the neighbourhood of $ SCDM $  at the time of evolution of the Universe without passing through $ \Lambda CDM $ and $ SCDM $converge to $ SS $, the steady state model of the Universe which suggest the steady state behavior of dark energy model in late times.

\item In section VI, Figs. 9 and 10. represent the likelihood contours for $ n $ and $ H_{0} $ with $ 1\sigma $, $ 2\sigma $ and $3\sigma $ errors in the $ n $-$ H_{0} $ plane. The constrained values of $ n $ are obtained $ 1.5176 $, $ 1.5907 $, $ 1.5009 $, $ 1.5396 $ and $ 1.5060 $ according to the Hubble $H(z)$, $ SNeIa $, $ BAO $, $H(z)+SNeIa$ and $H(z)+SNeIa+BAO$ data sets for which the corresponding values of $ H_0 $ are obtained as $ 68.0000 $, $ 66.3402 $, $ 66.4138 $, $ 65.6148 $ and $ 63.0579 $ respectively (see Table IV). Here, we observe that both constrained values of $n$ and $H_0$ are best fitted values according to these datasets.
 
\item In section VII, we have investigated some other types of the cosmological tests such as lookback time, proper distance, angular diameter distance and luminosity distance and it is found that the resulting outcomes of these kinematics test are consistent with current observations.
\end{itemize}

\vskip0.2in 
\noindent \textbf{Acknowledgements} R. Nagpal and J. K. Singh express their thanks to the Dept. of Mathematical Sciences, University of Zululand, S.A. for financial support and hospitality to complete the work. The author JKS also express his thanks to CTP, Jamia Millia Islamia, New Delhi, India for some fruitful discussions with Prof. M. Sami and Prof. S. G. Ghosh.

\end{document}